\documentclass[manuscript]{aastex}

\usepackage{emulateapj5}
\submitted{Astronomical Journal, in press (April 2001)}

\received{}
\revised{}
\accepted{}
\cpright{type}{year}
\journalid{VOL}{JOURNAL DATE}

\def\etal{{\em et~al.\ }}

\def\spose#1{\hbox to 0pt{#1\hss}}
\def\lta{\mathrel{\spose{\lower 3pt\hbox{$\mathchar"218$}}
     \raise 2.0pt\hbox{$\mathchar"13C$}}}
\def\gta{\mathrel{\spose{\lower 3pt\hbox{$\mathchar"218$}}
     \raise 2.0pt\hbox{$\mathchar"13E$}}}

\def\=#1{\overline{#1}}

\def\kms{{\rm\,km\,s^{-1}}}

\def\pc{{\rm\,pc}}
\def\kpc{{\rm\,kpc}}
\def\mpc{{\rm\,Mpc}}

\def\msun{{\rm\,M_\odot}}

\def\vcmax{v_c^{\rm max}}
\def\vcrmax{v_c(R_{\rm max})}

\begin{document}

\newlength{\figwidth}
\setlength{\figwidth}{0.5\hsize}

\title{Dynamical family properties and dark halo scaling relations 
of giant elliptical galaxies}
\author{Ortwin Gerhard, Andi Kronawitter}
\affil{Astronomisches Institut, Universit\"at Basel, Venusstrasse 7,
        CH-4102 Binningen, Switzerland; gerhard@astro.unibas.ch, krona@unibas.ch}
\author{R.P.~Saglia, Ralf Bender}
\affil{Institut f\"ur Astronomie und Astrophysik, Scheinerstra\ss e 1,
    D-81679 Munich, Germany; saglia@usm.uni-muenchen.de, bender@usm.uni-muenchen.de}

\shorttitle{Dynamical family properties of giant elliptical galaxies}
\shortauthors{O.E.~Gerhard, A.~Kronawitter, R.P.~Saglia, R.~Bender}



\begin{abstract}

Based on a uniform dynamical analysis of the line-profile shapes of 21
mostly luminous, slowly rotating, and nearly round
elliptical galaxies, we have investigated the dynamical family
relations and dark halo properties of ellipticals. Our results
include: (i) The circular velocity curves (CVCs) of elliptical
galaxies are flat to within $\simeq 10\%$ for $R\gta 0.2R_e$. (ii)
Most ellipticals are moderately radially anisotropic; their dynamical
structure is surprisingly uniform. (iii) Elliptical galaxies follow a
Tully-Fisher (TF) relation with marginally shallower slope than spiral
galaxies, and $v_c^{\rm max}\simeq 300\kms$ for an $L_B^\ast$ galaxy. 
At given circular velocity, they are $\sim$ 1 mag fainter in
B and $\sim$ 0.6 mag in R, and appear to have slightly lower
baryonic mass than spirals, even for
the maximum $M/L_B$ allowed by the kinematics.  (iv) The luminosity
dependence of $M/L_B$ indicated by the tilt of the Fundamental Plane
(FP) is confirmed. The tilt of the FP is not caused by dynamical or
photometric non-homology, although the
latter might influence the slope of $M/L$ versus $L$. It can also not
be due only to an
increasing dark matter fraction with $L$ for the range of IMF currently
discussed.  It is, however, consistent
with stellar population models based on published metallicities and
ages. The main driver is therefore probably
metallicity, and a secondary population effect is needed to explain
the K-band tilt. (v) These results make it likely that elliptical
galaxies have nearly maximal $M/L_B$ (minimal halos).
(vi) Despite the uniformly flat CVCs, there is a
spread in the luminous to dark matter ratio and in cumulative
$M/L_B(r)$. Some galaxies have no indication for dark matter within
$2R_e$, whereas for others we obtain local $M/L_B$s of 20-30
at $2R_e$. (vii)
In models with maximum stellar mass, the dark matter contributes $\sim
10-40\%$ of the mass within $R_e$.
Equal interior mass of dark and
luminous matter is predicted at $\sim 2-4R_e$.  (viii) Even in these maximum
stellar mass models, the halo core densities and phase-space densities
are at least $\sim 25$ times larger and the halo core radii $\sim 4$ times
smaller than in spiral galaxies of the same circular velocity. The
increase in $M/L$ sets in at $\sim 10$ times larger acceleration than in
spirals.  This could imply that elliptical galaxy halos collapsed at high
redshift or that some of the dark matter in ellipticals might
be baryonic.

\end{abstract}


\section{Introduction}

Hierarchical theories of galaxy formation predict that elliptical
galaxies should be surrounded by dark matter halos. The study of these
halos has been difficult, however, because of the lack of suitable and
easily interpreted tracers such as the HI rotation curves in spiral
galaxies. Recent work in several fields, however, leaves no doubt
about the existence of dark matter in ellipticals: X-ray data on their
hot gas athmospheres implies that dark halos in ellipticals are
ubiquitous and that the mass-to-light ratios are $M/L\sim 100$ on
scales of $\sim 100 \kpc$ (Mushotzky \etal \cite{Mush94}, Matsushita
\etal \cite{Metal98}, Loewenstein \& White \cite{LW99}). Gravitational
lensing studies (Kochanek \cite{Ko95}, Keeton, Kochanek \& Falco
\cite{KKF98}, Griffiths \etal \cite{Gri98}) show evidence for large
$M/L$ in lens elliptical galaxies, and stellar-dynamical studies based
on absorption line profile shapes have given strong constraints on the
mass distributions to $\sim 2R_e$ (Rix \etal \cite{Retal97}, Gerhard
\etal \cite{G+98}, Saglia \etal \cite{Sagl+00}), with a small to
moderate dark matter fraction inside $2R_e$.

Despite of this progress, the detailed mass distributions in
elliptical galaxies and their variations with luminosity remain
largely unknown.  Has the luminous matter segregated dissipatively in
the halo potential? Is there a ``conspiracy'' between luminous and
dark matter to produce a flat rotation curve, like in spiral galaxies?
How do the mass--to--light ratio, the slope of the circular velocity
curve, or the orbital anisotropy scale with luminosity? Is the tilt of
the Fundamental Plane simply related to a variation of the dynamical
$M/L$ with $L$?  Do elliptical galaxies follow a Tully-Fisher
relation? How do the scale radii and densities of elliptical galaxy
halos compare to those of spiral galaxies? 

The purpose of this paper is to address some of these questions on the
basis of a new dynamical study by Kronawitter \etal (\cite{K+00},
hereafter K+2000), who analyzed the dynamical structure and mass
distribution for a sample of 21 bright elliptical galaxies. Continuing
from previous work by our group (Gerhard \etal \cite{G+98}, Saglia
\etal \cite{Sagl+00}), these authors modelled the line profile shapes
of in total 17 E0/E1 and 4 E2 galaxies for which kinematic data
including line profile information were available from their own
observations or from the literature. The dynamical structure of these
galaxies turned out to be remarkably uniform. Most galaxies require
moderate radial anisotropy in their main bodies (at $\sim 0.5
R_e$). Their circular velocity curves are all consistent with being
flat outside $\simeq 0.2 R_e$.  The $M/L$ ratio profiles begin to rise
at around $0.5-2R_e$ and are consistent with X-ray and other data
where available, although from the kinematic data alone constant $M/L$
models can only be ruled out at $95\%$ confidence in a few galaxies.

This sample provides a new and much improved basis for investigating
the dynamical family properties of elliptical galaxies, which is the
subject of the present study. In Section 2, we analyze the
unexpectedly uniform dynamical structure of these elliptical galaxies.
In Section 3, we investigate the dependence on luminosity, discussing
the Faber-Jackson, Tully-Fisher and Fundamental Plane relations. In
Section 4, we relate the dynamical mass-to-light ratios to the stellar
population properties. In Section 5 we discuss the structure of the
dark halos of these ellipticals.  Our conclusions are summarized in
Section 6.

\section{Dynamical structure}
\label{secstructure}

The elliptical galaxies analyzed by Kronawitter \etal (\cite{K+00}, K+2000)
divide in two subsamples, one with new extended kinematic data,
reaching typically to $\sim 2R_e$ (`EK sample', the data are from
Kronawitter \etal and from several other sources referenced there),
and one with older and less extended kinematic measurements (`BSG
sample'; this is a subsample from Bender, Saglia \& Gerhard \cite{BSG94}).
Based on these data and mostly published photometry, K+2000
constructed non-parametric spherical models from which circular
velocity curves, radial profiles of mass-to-light ratio, and
anisotropy profiles for these galaxies were derived, including
confidence ranges.

The galaxies were selected to rotate slowly if at all and to be as
round as possible on the sky. They are luminous elliptical galaxies
($M_B\simeq -21\pm2$)\footnote{Throughout this paper we use a Hubble
constant $H_0=65\kms\mpc^{-1}$ unless explicitly noted otherwise.}. The
expected mean intrinsic short-to-long axis ratio for such a sample of
luminous ellipticals is $\!<c/a\!>=0.79$. The mean systematic effects
arising from the use of spherical models and the possible presence of
small embedded face-on disks are small for the sample as a
whole, but may be non-negligible in individual cases (see K+2000, \S5.1).

\subsection{Circular velocity curves}
\label{seccvc}

\begin{figure*}[t]
\begin{center}
\resizebox{\figwidth}{!}{\includegraphics{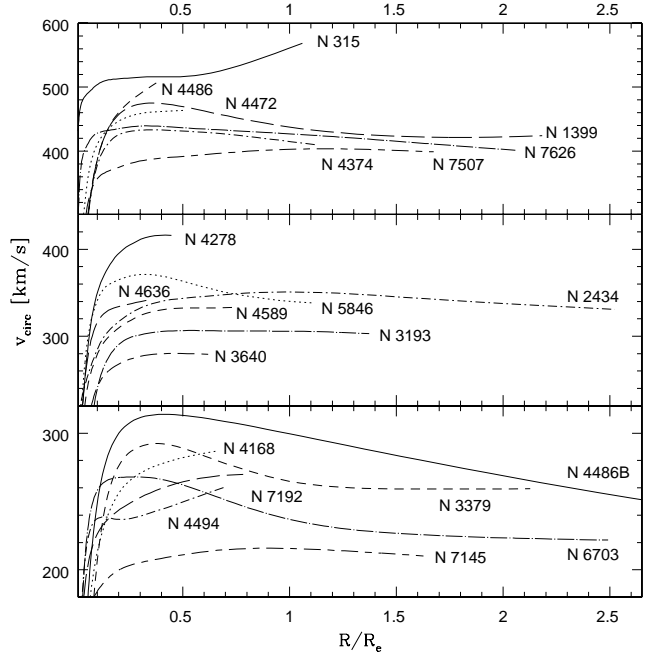}}
\caption[dsb]{\small The ``best model'' circular velocity curves of
all galaxies from the K+2000 sample plotted as a function of
radius scaled by the effective radius $R_e$. The panels are
roughly ordered by luminosity.}
\label{vcall}
\end{center}
\end{figure*}

\begin{figure*}[t]
\begin{center}
\resizebox{\figwidth}{!}{\includegraphics{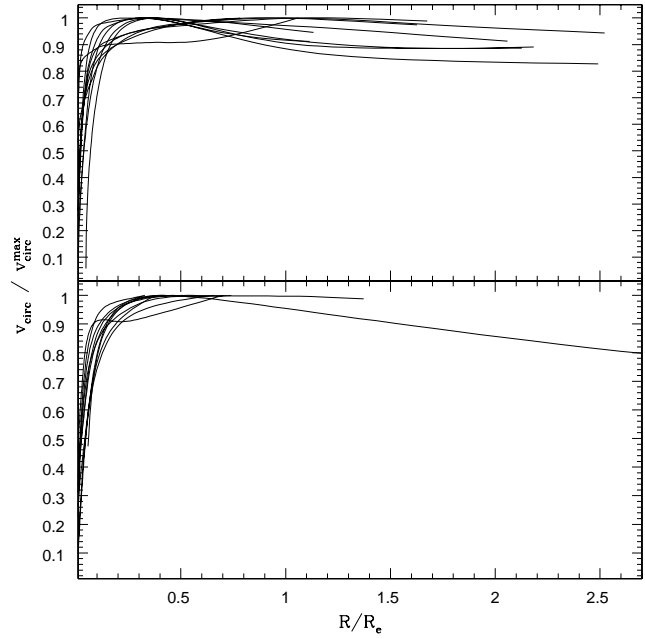}}
\caption[dsb]{\small Same circular velocity curves, normalized by the
maximum circular velocity. The upper panel now shows the galaxies from
the EK subsample of K+2000, the lower panel those from the BSG
subsample. The extended curve in the lower panel is for the compact
elliptical NGC 4486B.}
\label{vcnorm}
\end{center}
\end{figure*}

Circular velocity curves (CVCs) for all galaxies in the sample are
shown in Figure \ref{vcall}, in three bins roughly ordered by
luminosity.  CVCs normalized by the respective maximum circular
velocity are shown in Figure \ref{vcnorm} separately for the two
subsamples. The plotted curves correspond to the ``best'' models of
K+2000, which are taken from the central region of their 95\%
confidence interval for each galaxy, respectively. Based on dynamical
models near the boundaries of the confidence interval, the typical
uncertainty in the outermost circular velocity is $\pm$(10-15)\%. The
expected mean systematic error from flattening along the line-of-sight
is smaller; cf.~\S5.1 of K+2000.

The most striking result from these diagrams is that at the
$\simeq10\%$ level all CVCs are flat outside $R/R_e \simeq 0.2$.  This
result is most significant for the galaxies with the extended data,
while for many galaxies from the BSG sample the radial extent of the
data is insufficient to show clear trends. However, in cases where
X-ray data are available (NGC 4472, 4486, 4636) the mass profiles of
the ``best'' models approximately match those from the X-ray analysis
even for those galaxies (see K+2000).

\begin{figure*}[t]
\begin{center}
\resizebox{\figwidth}{!}{\includegraphics{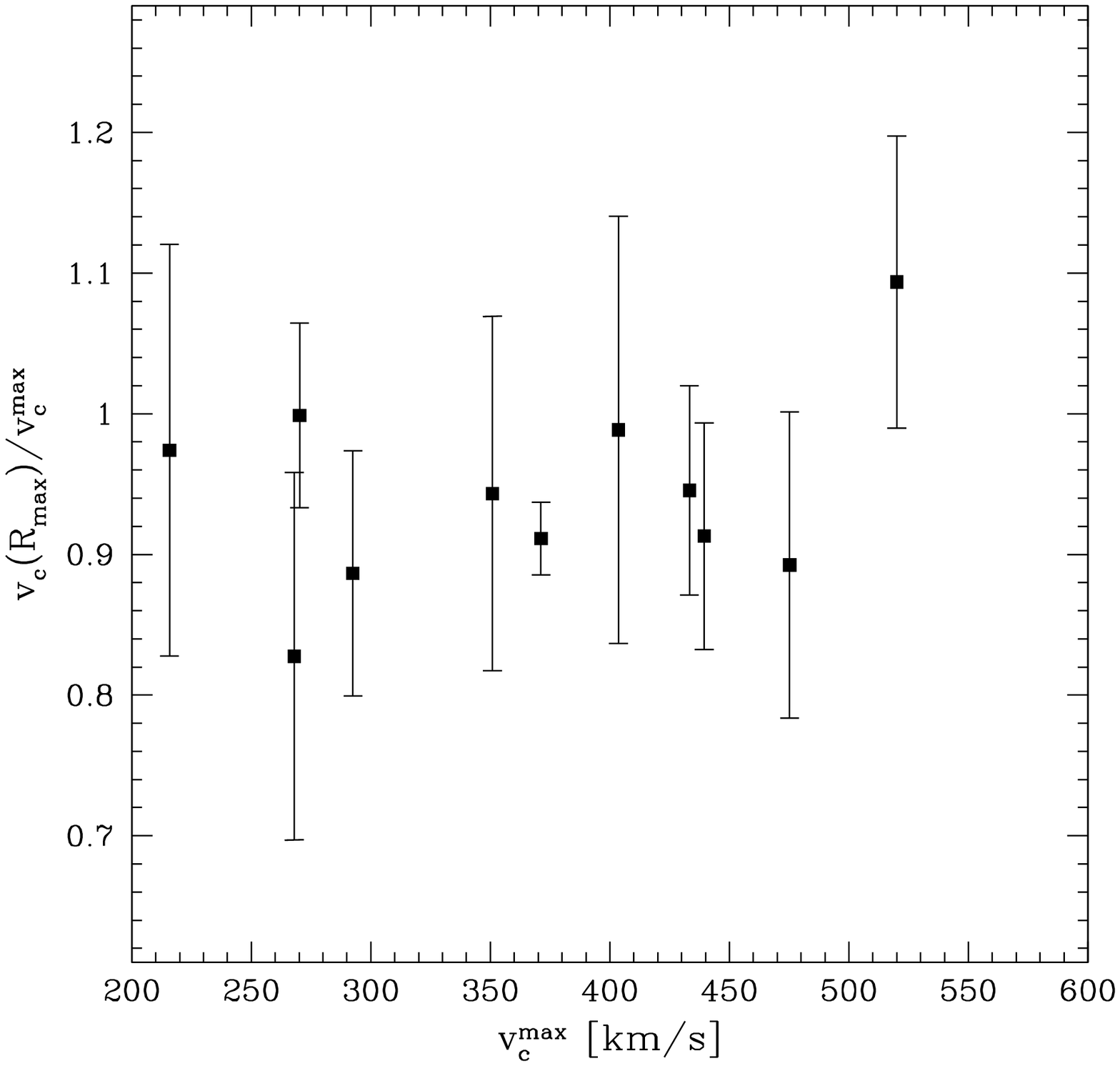}}
\caption[dsb]{\small The ratio $\vcrmax/\vcmax$ for all galaxies from
  the EK-sample, showing the gradient in the circular velocity curves. 
  The error-bars plotted
  correspond to the range of $\vcrmax$ from the 95\%-confidence models
  of K+2000.}
\label{vcerr}
\end{center}
\end{figure*}

This result is illustrated further by Figure \ref{vcerr}, which shows
the derived ratio $\vcrmax/\vcmax$ for all galaxies of the EK-sample.
Here $\vcrmax$ is the circular velocity at the radius of the last
kinematic data point, and $\vcmax$ is the maximum circular velocity in
the respective ``best'' model.  For NGC 315 $v_c(0.6R_e)$ was used
instead of $\vcmax$. The error-bars plotted correspond to the $95\%$
confidence range for $\vcrmax$, compared to which the uncertainty in
$\vcmax$ can be neglected.  Most of these galaxies have
$\vcrmax/\vcmax\simeq 0.9-1.0$ with a median at 0.94, and 95\%
confidence ranges  $\sim\pm 0.1$.

The galaxies in the EK sample appear to show a ``bimodal''
distribution of CVC shapes in Figs.~\ref{vcall}, \ref{vcnorm}: For one
group of galaxies the ``best'' model CVC has a peak near $0.3 R_e$ and
then falls slightly until it becomes flat at $\simeq 1 R_e$ (NGC 1399,
NGC 3379, NGC 5846 and NGC 6703, the latter has the largest drop). In
the other group the CVC rises rapidly until $\simeq 0.2 R_e$ and
reaches a peak at $\simeq 1 R_e$ after which it remains nearly flat
(NGC 2434, NGC 7145, NGC 7192, NGC 7507). NGC 4374 and NGC 7626 may be
cases from the first group where the flat part is not yet seen in
the data.  The rise of the CVC of NGC 315 seen near $R_{\rm max}$ in
Fig.~\ref{vcall} may not be real; a flat CVC appears also consistent
with the kinematic data (see discussion in K+2000).  The difference
between these CVC shapes is about a $\simeq 2\sigma$ result when
comparing the model confidence bands for EK galaxies of either type,
but this estimate does not include possible systematic effects from
the use of spherical models.  If the differences in CVC shapes are
real, they could reflect small variations in the degree of dissipation
and mass segregation of the baryonic component during the formation
process. In any case no clear trend with galaxy luminosity is seen.

\subsection{Anisotropy}
\label{secanisotropy}

\begin{figure*}[t]
\begin{center}
\resizebox{\figwidth}{!}{\includegraphics{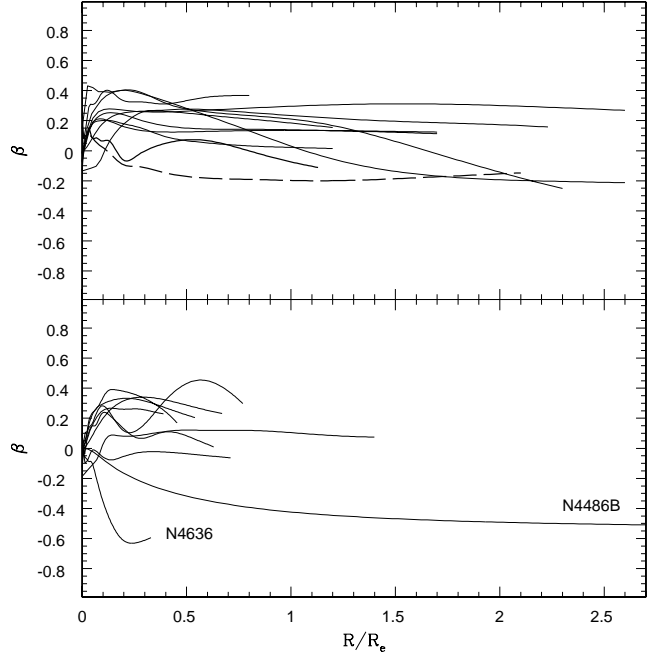}}
\caption[dsb]{\small Velocity anisotropy $\beta$ as a
function of radius. Upper panel: for the galaxies with extended data
(EK subsample), lower panel: for the galaxies from the BSG subsample.}
\label{bet}
\end{center}
\end{figure*}

Figure \ref{bet} shows the radial profiles of the anisotropy parameter
$\beta=1-\sigma_t^2/\sigma_r^2$ for all galaxies. Here $\sigma_r$ and
$\sigma_t$ are the intrinsic radial and one-dimensional tangential
velocity dispersions, and $\beta=1,0,-\infty$ for completely radial,
isotropic, and circular orbit distributions, respectively.  Because it
is a deprojected quantity, the anisotropy derived from our models is
considerably more uncertain than the total mass $M(r)$; see Gerhard
\etal (\cite{G+98}) for further discussion. The uncertainties are
particularly large near the outer boundary of the data, where the
range of $\beta$ values in models corresponding to the allowed
potentials is usually 0.2--0.5. For the inner profiles, the allowed range is
typically 0.1-0.2, sometimes 0.3. Nonetheless, Figure \ref{bet} shows
a clear trend in that almost all galaxies are radially anisotropic in
the inner (best-constrained) regions, with the peak of the anisotropy
often near 0.2 $R_e$ with values of $\beta = 0.2 \dots 0.4$. There are
two cases in the EK sample which are exceptional in that they are
consistent with isotropy over the whole radial range. These two are
NGC 315 (thick solid line), a cD galaxy, and NGC 7626 (thick long
dashed line), which is classified as E pec. Despite their less
extended and lower S/N data, the galaxies from the BSG sample show a
similar trend, again with two exceptions: NGC 4636, which we do not
consider a very reliable case, and NGC 4486B, which is exceptional in
almost all respects.

\begin{figure*}[t]
\begin{center}
\resizebox{\figwidth}{!}{\includegraphics{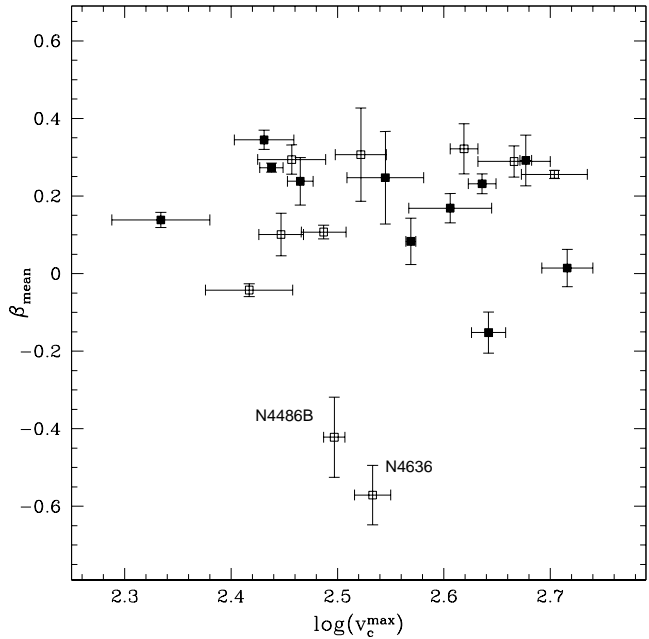}}
\caption[dsb]{\small Average velocity anisotropy $\beta$, calculated
over the range $0.1 R_e$ to $1 R_e$ or $R_{\rm max}$ if $R_{\rm max}
< R_e$, versus maximum circular velocity. Solid squares: galaxies with
extended data, EK subsample; open squares: galaxies from BSG
subsample; see text and K+2000.  The error-bars plotted for $\beta$
show the standard deviation from the mean in the radial range used in
the averaging.
}
\label{betmvc}
\end{center}
\end{figure*}

To get a more robust estimate of the anisotropy we have averaged the
anisotropy profile between 0.1 $R_e$ and min$(R_e, R_{\rm max})$,
where $R_{\rm max}$ is the radius of the outermost kinematic data
point. These averaged anisotropies are plotted in Figure \ref{betmvc}
versus the maximum circular velocity from Fig.~\ref{vcall}. In a few
cases this is the circular velocity at the outer boundary of the
modelled range.  Most galaxies with extended data show a clear maximum
of the CVC whereas the BSG galaxies have CVCs which may still rise
outwards. For these the true maximum probably lies outside the range
of the kinematic data.  The error-bars plotted for $\beta$ show the
standard deviation from the mean over the radial range used in the
averaging. The errors plotted for $v_c$ are equal to $\pm$ one half
the separation of the extreme models in the confidence interval, near
the radius where the best model has its maximum.  The figure shows
again that all galaxies but two (NGC 4486B and NGC 4636) have a mean
radial anisotropy in the range from $\beta\simeq 0$ to $\beta\simeq 
0.35$, and also that there is no dependence on circular velocity (or
luminosity).

Radial anisotropy has also been inferred from three-integral
axisymmetric models for several flattened elliptical galaxies: NGC
1600 (E3.5, Matthias \& Gerhard \cite{MG99}), NGC 2300 (E2, Kaeppeli
\cite{Ka99}), NGC 2320 (E3.5, Cretton, Rix \& de Zeeuw \cite{CRZ00},
NGC 3379 (E1, Gebhardt \etal \cite{Gebh+00}). This makes it unlikely
that our results are severely biased by the use of spherical dynamical
models. As discussed in K+2000, the mean intrinsic short-to-long axis
ratio for our sample of elliptical galaxies is $\!<c/a\!>=0.79$.
Because these are luminous galaxies, rotation will not contribute
substantially to any flattening, so except for the flattest galaxies
the bias introduced by face-on circular orbits will be small. On the
other hand, possibly embedded face-on disks are likely to be less than
$0.1-0.3R_e$ in size on statistical grounds (Mehlert \etal
\cite{Meetal98}). In three galaxies of our sample, such disks are
known and extend to 6, 7, and 8 arcsec in NGC 4472, NGC 4494, NGC
7626, respectively. In these very inner regions, these disks might
cause the anisotropy to be overestimated by $\simeq 0.2$, but they
will not affect the globally averaged results significantly.

\subsection{$v_c^{\rm max}$--$\sigma_{\rm 0.1}$}
\label{secvcsigma}

\begin{figure*}[t]
\begin{center}
\resizebox{\figwidth}{!}{\includegraphics{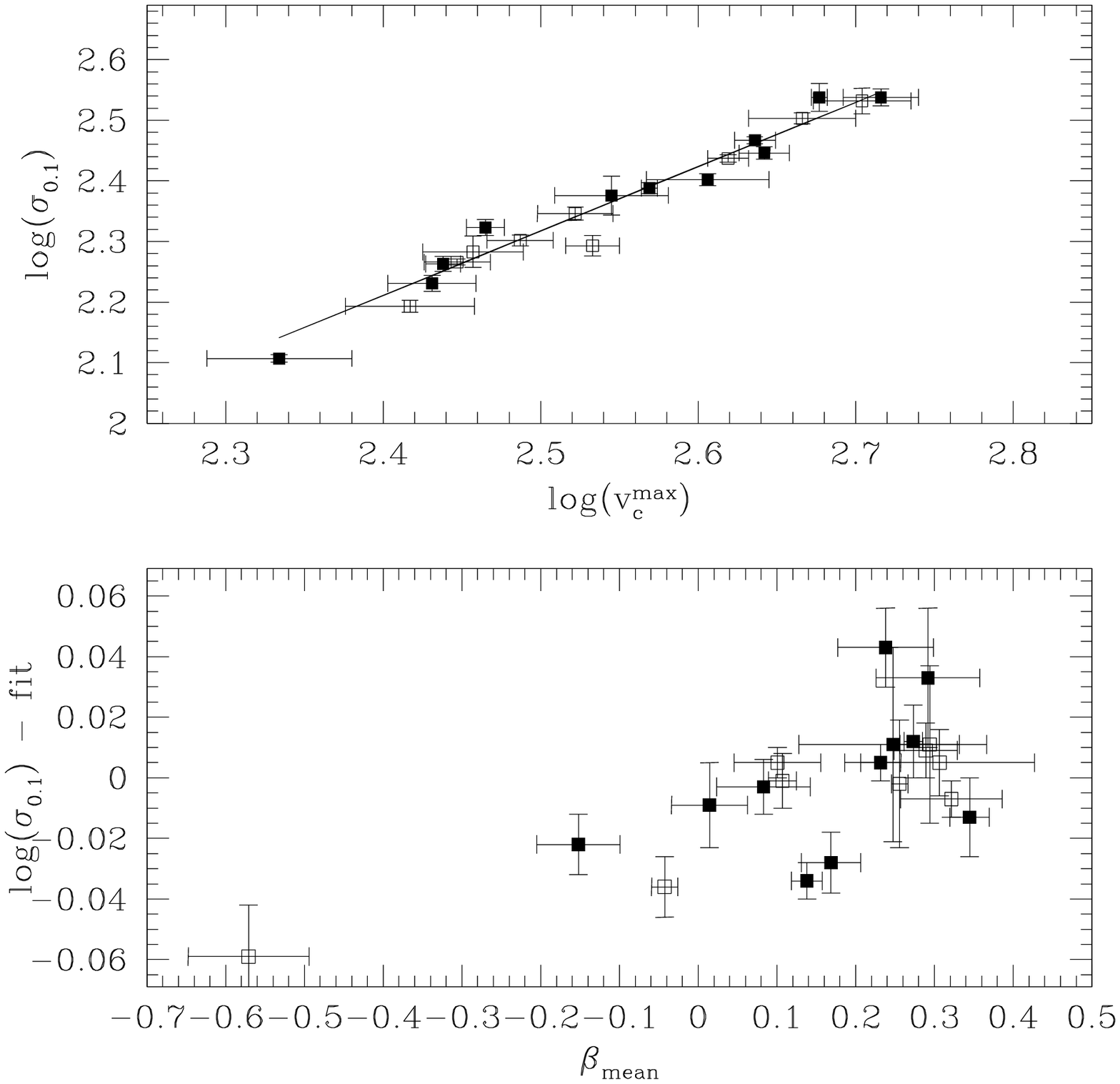}}
\caption[dsb]{\small Top: the correlation between the averaged central
velocity dispersion and the inferred maximum circular
velocity. Bottom: the residuals from this correlation plotted versus
the mean orbital anisotropy defined in Section \ref{secanisotropy}.
Symbols as in Fig.~\ref{betmvc}.}
\label{vcsig}
\end{center}
\end{figure*}

We have found that both the circular velocity curves and the
anisotropy profiles of elliptical galaxies are surprisingly similar.
In zeroth order the CVCs can thus be characterized by two scaling
constants, the effective radius and a velocity scale.  In
Fig.~\ref{betmvc} we have used the maximum circular velocity; but in
fact it should not matter which velocity is used to set the scale. In
particular, we would expect that a suitably defined central velocity
dispersion could equally be used.  A little care is needed, however,
since the measured central velocity dispersion may be influenced by
the gravitational field of a central black hole and by the resolution
of the kinematic data. We therefore use an average central velocity
dispersion $\sigma_{0.1}$, defined as the square root of the average
of all measured $\sigma^2_i\equiv\sigma^2(R_i)$ inside $0.1 R_e$ or
3'', whichever is the larger of the two radii. This would be the inner
luminosity-averaged rms velocity dispersion if the surface brightness
profile were exactly proportional to $R^{-1}$.

Figure \ref{vcsig} shows a plot of $\sigma_{0.1}$ versus the maximum
circular velocity for all elliptical galaxies in our sample but NGC
4486B.  The error-bars for $v_c^{\rm max}$ are taken from the allowed
model range as before. For the error-bars in the $\sigma_{0.1}$ we
have taken the larger of
\begin{equation}
{\Delta\sigma_{0.1} \over \sigma_{0.1}} =
  {[\sum_i(\sigma^2_i-\sigma^2_{0.1})^2]^{1\over 2}
        \over
   2 \sqrt{N-1} \sigma^2_{0.1} }
\end{equation}
and a weighted observational error in $R\le 0.1 R_e$,
$\Delta\sigma_{\rm obs}\equiv \left( \sum_i \Delta\sigma_i^{-2} \right)^{-1/2}$.
For some galaxies the actual error of $\sigma_{0.1}$ may be smaller than
this conservative estimate from the standard deviation of the
$\sigma_i^2$, but we have used this because several galaxies have
velocity dispersion gradients in the central $0.1 R_e$.
Fig.~\ref{vcsig} shows a very good correlation; the slope is $1.062\pm
0.058$ for a reduced $\chi^2=0.87$, as determined by the routine {\em
  fitexy} of Press \etal (\cite{Petal92}). This routine fits a
straight line to the datapoints by minimizing a $\chi^2$ function
which involves the errors in both the x- and y-variables.  Thus the
fitted relation is completely consistent with linear; explicitly,
\begin{equation}
\sigma_{0.1}=0.66 v_c^{\rm max}.
\end{equation}

The lower panel of Figure \ref{vcsig} shows that the residuals from
this relation are correlated with the mean orbital anisotropy defined
in \S\ref{secanisotropy} (the correlation would be somewhat stronger
had we plotted it in terms of the maximum anisotropy).  Differences in
dynamical structure thus cause some scatter in the relation between
$\sigma_{0.1}$ and $v_c^{\rm max}$ but their influence on this and
other global correlations is small. In this sense the dynamical
structure of these ellipticals is indeed very uniform.


\section{Scaling relations}
\label{secscalings}

\begin{figure*}
\begin{center}
\resizebox{\figwidth}{!}{\includegraphics{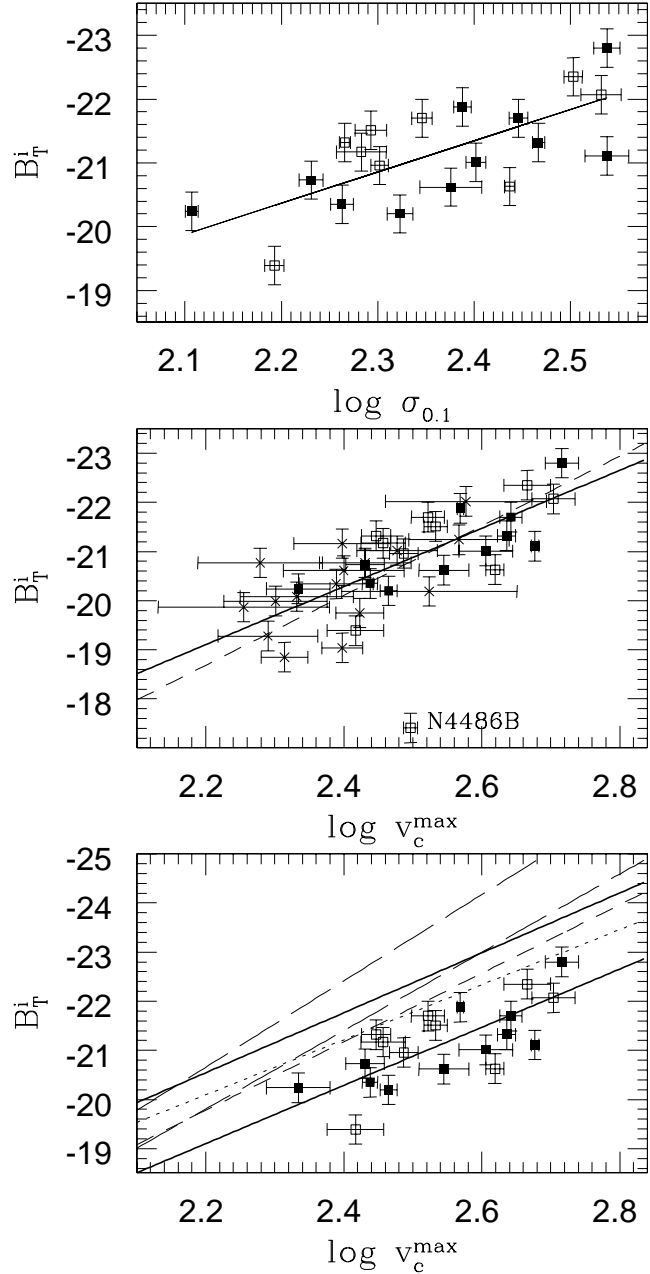}}
\caption[dsb]{\small 
Top panel: Faber-Jackson relation for our sample of elliptical
galaxies. Full squares and open squares show galaxies from the EK and
BSG subsamples of K+2000.  $\sigma_{0.1}$ is the averaged velocity
dispersion inside $0.1R_e$ (see text); this is not available for NGC
4486B. The full line shows a least-square fit.  Middle panel:
Tully-Fisher relation for the elliptical galaxies from K+2000 (same
point styles) and S0s from Neistein \etal \cite{Netal99} (stars).
Ellipticals and S0s form a smooth sequence; the slope of the
elliptical galaxies alone (full line, fit excludes NGC 4486B) is marginally
shallower than the slope for the combined sample (dashed line).  Lower
panel: Comparison of elliptical and spiral galaxy Tully-Fisher
relations. Ellipticals: Data points and least-square fit line repeat
the B-band relation. The upper heavy line shows this relation shifted
to the R-band, using the colour-magnitude relation given in the
text. Spirals: Cepheid-calibrated B-band Tully-Fisher relations from
Federspiel \etal (\cite{FTS98}, short-dashed) and Sakai \etal (\cite{Sakai+00}, lower
long-dashed), Cepheid-calibrated R-band relation from Sakai \etal
(\cite{Sakai+00}, upper long-dashed), and r-band relation from Courteau \& Rix
(\cite{CouRix99}, dotted) shifted to the B-band using the colour-magnitude 
relation for their spirals.
}
\label{tfFj}
\end{center}
\end{figure*}

\subsection{Faber-Jackson relation}

The well-known relation between the total magnitude and central
velocity dispersion for elliptical galaxies (Faber and Jackson
\cite{FJ76}) is shown in the top panel of Figure \ref{tfFj} for the
K+2000 sample of ellipticals, with $\sigma_{0.1}$ in place of the
central dispersion.  In this plot and all subsequent similar plots
B$^i_T$ is the twice the total integrated B-band absolute magnitude
within $R_e$ (column 7 of Table 4 in K+2000). The errors in these
B$^i_T$ magnitudes are set to 0.3 mag. This accounts for observational
errors and uncertainties in the distances. The latter are estimated as
approximately 0.25 mag from the intrinsic scatter of the Dn-$\sigma$
relation, the depth of clusters (in case of group/cluster distances),
and a comparison of various distance determinations by Tonry \etal
(\cite{Tetal97}). The fitted slope of the Faber-Jackson relation for
our ellipticals (again using the errors in both variables) is
$-4.89\pm1.12$, i.e., $L_B\propto (\sigma_{0.1})^{1.96\pm 0.45}$.  The
fit excludes NGC 4486B; this galaxy is a close companion of M87 and
its low luminosity for its high circular velocity suggests that the
galaxy may be tidally disturbed.  The uncertainty in the slope is
determined with rescaled errors such that the reduced $\chi^2=1.0$.
The shallow slope is consistent within our errors with the slope
determined from the data of Faber \etal (1989), which give $L_B\propto
\sigma^{2.61\pm 0.08}$ using their R-distances. For comparison, the
K-band slope is significantly steeper (Pahre \etal \cite{Petal98}).

\subsection{Tully-Fisher relation}
\label{secTF}

\begin{figure*}[t]
\begin{center}
\resizebox{\figwidth}{!}{\includegraphics{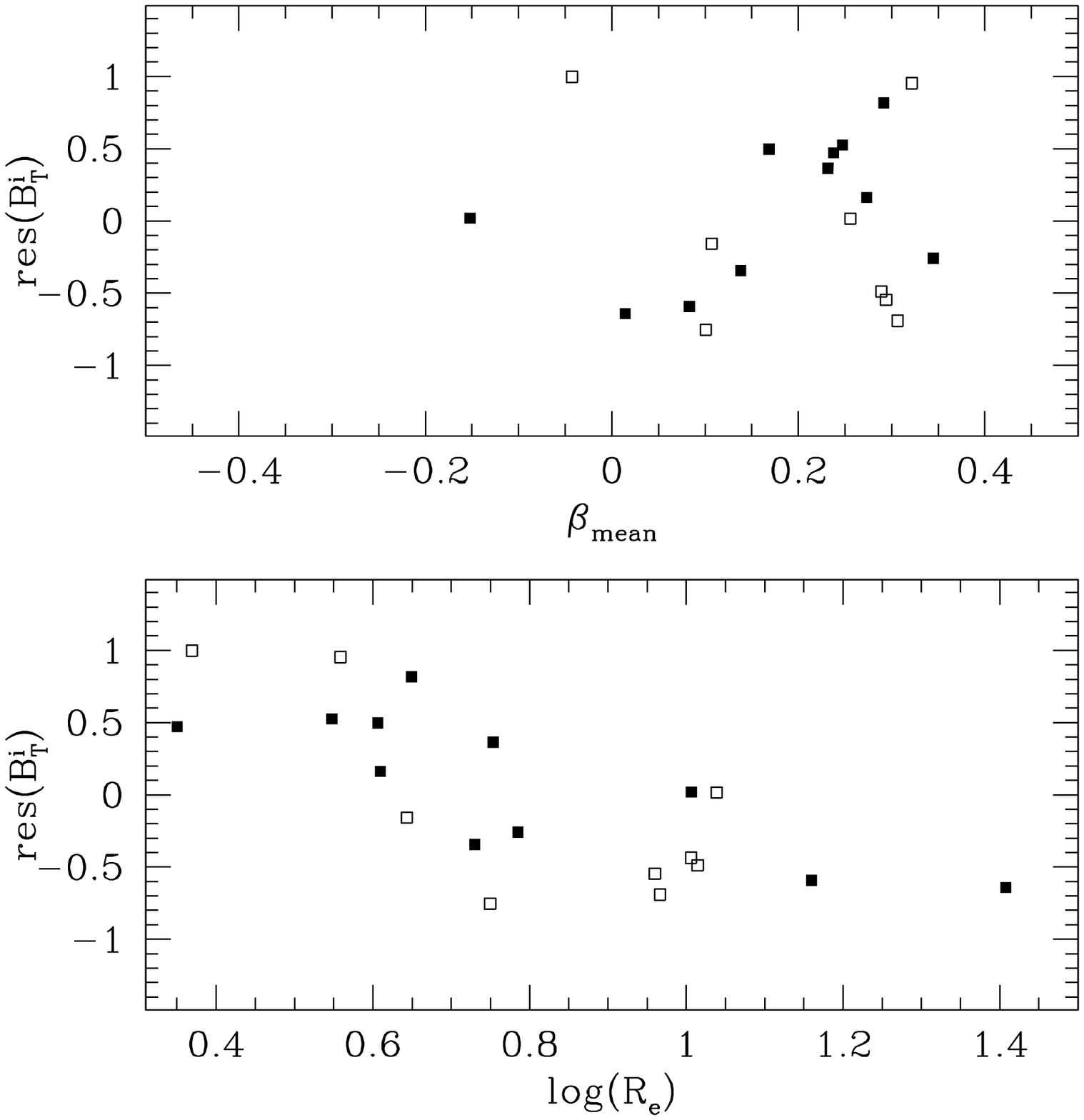}}
\caption[dsb]{\small 
Top panel: The residuals in B$_T$ from the
Tully-Fisher relation for the elliptical galaxies versus their
mean anisotropy $\beta_{\rm mean}$, excluding NGC 4486B.
Lower panel: The residuals in B$_T$ from the
elliptical galaxy Tully-Fisher relation versus effective radius $Re$.
}
\label{tfresds}
\end{center}
\end{figure*}

The relation between the circular velocity and total luminosity, known
as Tully-Fisher (\cite{TF77}, TF) relation, is observationally well--established
for spiral galaxies.  From the K+2000 dynamical analysis, we now know
circular velocities also for elliptical galaxies, and can thus
investigate whether ellipticals also follow a TF relation.  We use the
$v_c^{\rm max}$ variable and errors from \S\ref{secstructure} and
integrated total B$^i_T$ as before, for the 21 galaxies analyzed in
K+2000.  These data points are shown in the middle and lower panels of
Figure \ref{tfFj}.  From a least-square fit in both variables we
obtain a B-band TF-slope of $-5.92 \pm 1.21$, corresponding to
\begin{equation}
L_B\propto (v_c^{\rm max})^{2.37\pm 0.48},
\end{equation}
again excluding NGC 4486B and using rescaled errors to estimate the
uncertainty. The derived slope is
consistent with the Faber-Jackson relation and the $v_c^{\rm
max}$--$\sigma_{\rm 0.1}$ relation discussed in \S\ref{secvcsigma}.
The zero-point of the relation is given by
\begin{equation}\label{eqtfnorm}
  v_c^{\rm max} = 493 L_{11}^{0.42} \kms,
\end{equation}
where $L_{11}=L_B/10^{11}h_{0.65}^{-2} L_{\odot,B}$ and $h_{0.65}=H_0
/ 65 \kms\mpc^{-1}$; this implies $v_c^{{\rm max},\ast}\simeq 303\kms$
for an $L^\ast$--elliptical galaxy, using a corrected
$M_B^\ast(B_T^0)=-20.8$ ($L_B^\ast(B_T^0)=10^{10.5}L_{B,\odot}$)
for $H_0=65 \kms\mpc^{-1}$ from Fukugita \&
Turner (\cite{FT91}).  The corresponding $\sigma_{0.1}^\ast\simeq
195\kms$.

We have also included the S0s from Neistein \etal (\cite{Netal99}) in
the TF plot. We took the circular velocities corrected for asymmetric
drift from their Table 1, column 13, but used the B-band luminosities
from the RC3 (de Vaucouleurs \etal \cite{dVetal91}) rescaled to our
distance scale (using distances from Faber \etal (\cite{Fetal89})
where available: NGC 584, 1052, 2768, 3115, 4649, and for the remaining
S0 galaxies redshifts with respect to the CMB frame and
$H_0=65$ km/s/Mpc).  The S0's join smoothly with the ellipticals in
the TF plot.  The fitted slope for both samples together is slightly
steeper ($-7.09\pm0.91$) than for the ellipticals alone, but the two
slopes are consistent within their errors.

The lower panel of Fig.~\ref{tfFj} shows the comparison of the
elliptical and spiral galaxy TF relations. The data points and the
heavy least-square fit line repeat the B-band elliptical galaxy
relation. The two lower dashed lines show the Cepheid-calibrated
B-band Tully-Fisher relations for spiral galaxies from Federspiel
\etal (\cite{FTS98}) and Sakai \etal (\cite{Sakai+00}), which are in
mutual agreement to a few tenths of a magnitude over the range of
interest here. The TF relations of these authors are given in terms of
HI linewidth measured at 20\% peak flux and were converted to
$v_c^{\rm max}$ by using Fig.~18 of Rubin \etal (\cite{Retal99}),
which shows that the inclination-corrected $W_{20}=2.0 v_c^{\rm max}$
to within the errors. The dotted line shows the r-band relation from
Courteau \& Rix (\cite{CouRix99}), converted to the B-band using the
colour magnitude relation they give for their sample of spirals, and
converted to our $H_0=65\kms\mpc^{-1}$ distance scale. Their velocities
$v_{2.2}$ were used directly as $v_c^{\rm max}$. The average of the
spiral galaxy slopes is somewhat steeper than the elliptical galaxy
relation, but given the large scatter in the latter and between the
spiral galaxy slopes, this difference is marginal.

\begin{figure*}[t]
\begin{center}
\resizebox{\figwidth}{!}{\includegraphics{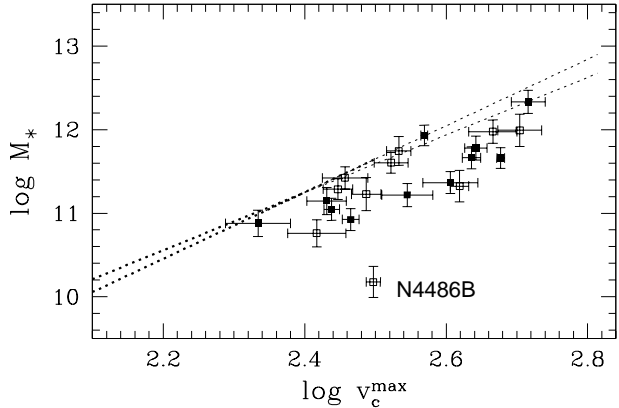}}
\caption[dsb]{\small Luminous mass in stars versus maximum circular velocity
for our sample of elliptical galaxies. The dotted lines show the baryonic mass
rotation velocity relations for spiral galaxies from McGaugh \etal 
(\cite{McG00}, steeper) and Bell \& de Jong (\cite{BdJ00}, shallower). The heavy
parts of these lines cover the range of the fitted spiral galaxy data.}
\label{tfbary}
\end{center}
\end{figure*}

Federspiel \etal (\cite{FTS98}) applied their calibration to a complete sample
of Virgo cluster spiral galaxies to derive a mean cluster distance
modulus of (m-M)$_0 = 31.58\pm 0.24$ mag.  According to Faber \etal
(\cite{Fetal89}), the Virgo cluster center is at a distance of $v=1333\kms$. By
using relative distances from Faber \etal (\cite{Fetal89}) (the R values of
their Tables 3, 4) and a Hubble constant of $H_0=65\kms\mpc^{-1}$ to
obtain the absolute magnitudes B$^i_T$ for our elliptical galaxies, we
have thus implicitly assumed a Virgo cluster distance modulus of
m-M$_0 = 5\log(1333/65) +25=31.56$ mag, which is identical to the
spiral galaxy mean Virgo distance modulus of Federspiel \etal (\cite{FTS98}).
Thus the zero points of the Cepheid-calibrated TF relations are
directly comparable to our elliptical galaxy TF. In fact, this
comparison need not make any assumption about absolute distances; the
only assumption made is that the centroid of the Virgo cluster spiral
galaxy sample coincides with the Virgo cluster ellipticals as given by
Faber \etal (\cite{Fetal89}); the relative distances between the ellipticals in
our sample are then also fixed. Therefore, the offset between
elliptical and spiral galaxies in Fig.~\ref{tfFj} can be regarded as
one in {\sl apparent magnitude}.  Its value depends only on having
used the correct relative distance of the two systems.

Taking an average over the different spiral galaxy TF relations in the
plot, we conclude that, at a given circular velocity, elliptical
galaxies are about 1 mag fainter in B than spiral galaxies.  Put
differently, at given luminosity elliptical galaxies have higher
circular velocities than spirals by about 0.2 dex. This difference
decreases slightly if instead of maximum circular velocities we use
the halo velocities of our models. To see whether this difference is
smaller in the redder R-band, where elliptical galaxies should be
relatively brighter, we have plotted in the lower panel of
Fig.~\ref{tfFj} the R-band TF relation expected from the B-band fit
and the colour-magnitude relation from the EFAR sample (Saglia \etal
\cite{Sagl97a}), B-R$=-0.030 ({\rm R} +22.5) + 1.50$ (upper full
line), to be compared to the R-band Cepheid-calibrated relation from Sakai
\etal (\cite{Sakai+00}), plotted as upper long-dashed line. With
respect to the Sakai \etal relations in R and B, the offset is indeed
$\sim 0.4$ mag smaller in R than in B at $\log v_c^{{\rm max},\ast}$,
but note the different slopes.

In Figure \ref{tfresds} we show the residuals from the elliptical
galaxy TF relation versus the mean velocity anisotropy $\beta_{\rm
mean}$ and the effective radius. There is no obvious correlation with
$\beta_{\rm mean}$.  On the other hand, the plot against $R_e$ does
show a correlation, which reflects the existence of the fundamental
plane (see next section).

In summary, elliptical galaxies follow a Tully-Fisher relation, with a
B-band slope that is marginally shallower than the slope for spiral
galaxies, and zero point such that an $L^\ast$--elliptical has a
circular velocity $v_c^{{\rm max},\ast}\simeq 300\kms$. At given
circular velocity, elliptical galaxies are about 1 mag fainter in B
than spiral galaxies.

\subsection{Baryonic Tully-Fisher relation}
\label{sectfbary}

\begin{figure*}[t]
\begin{center}
\resizebox{\figwidth}{!}{\includegraphics{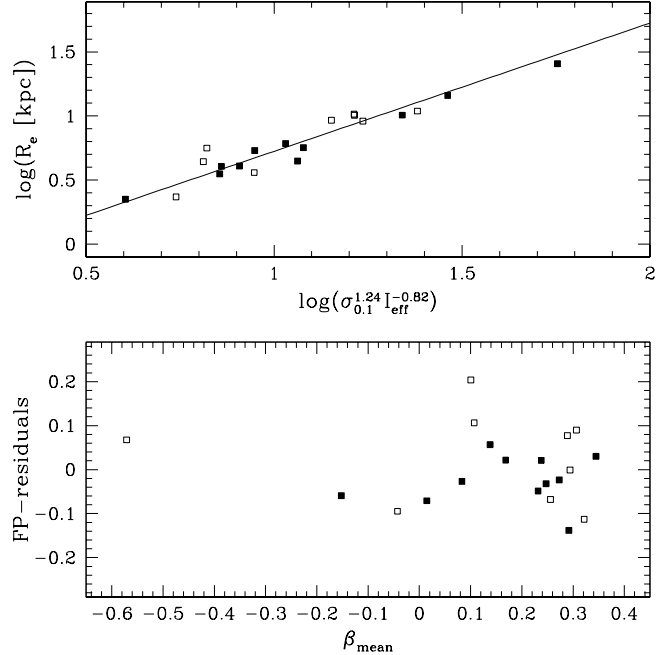}}         
\caption[dsb]{\small Upper panel: B-band Fundamental Plane for the
sample of elliptical galaxies from K+2000. The parametrization of the
FP is taken from J{\o}rgensen \etal (\cite{Joerg96}). The line shows the best fit
with fixed slope 1.0. 
Lower panel: residuals of the fit plotted against mean anisotropy
$\beta$. Symbols as in Fig.~\ref{betmvc}.}
\label{fpfitres}
\end{center}
\end{figure*}

Since the stellar mass-to-light ratios of elliptical galaxies vary
significantly with luminosity (see \S\ref{seclml}), it may be more
revealing to plot luminous mass against circular velocity than
luminosity. The mass of the X-ray emitting gas in ellipticals is only
a few percent even for luminous galaxies (Forman, Jones \& Tucker
\cite{Foretal85}, Sarazin \cite{Sar97}).  The stellar mass is thus
nearly equal to the total baryonic mass.  Fig.~\ref{tfbary} shows
stellar mass $M_\ast = M/L_{\rm central} \times L_B$ versus maximum
circular velocity for our sample of elliptical galaxies, where
$M/L_{\rm central}$ is the maximum $M/L$ allowed by the kinematic
data. The steeper dotted line in the diagram is the best fit spiral
galaxy baryonic Tully-Fisher relation relation from McGaugh \etal
(\cite{McG00}), including stellar and gaseous mass. Their stellar
masses were determined from red and NIR luminosities, using $M/L$'s
based on population models (constant star formation rate, Salpeter
IMF, such that $M/L_K=0.8 \msun/L_{\odot,K}$, $M/L_B=1.4
\msun/L_{\odot,B}$). According to McGaugh \etal (\cite{McG00}) these
model $M/L$ are consistent with maximum disk fits for the bright
galaxies.  The shallower dotted line is the spiral galaxy line from de
Bell \& de Jong (\cite{BdJ00}) . This is based on luminosities in several
passbands and M/L ratios determined from evolution models, which use
an IMF containing fewer low-mass stars than a Salpeter IMF, as
suggested by recent observations in the Galaxy (e.g., Gould, Bahcall
\& Flynn 1997, Holtzman \etal \cite{Hetal98}) and so that maximum disk
M/Ls are not exceeded. Both spiral galaxy relations apparently predict
very similar baryonic masses for the circular velocities of interest
here. The thin parts of both lines are extrapolated
from the thick parts with spiral galaxy data.

In Fig.~\ref{tfbary} elliptical galaxies fall below the spiral galaxy
lines by a factor of about two in the mean.  Note that the offset to
the spiral galaxies cannot be explained by luminosity errors for our
ellipticals alone (Saglia \etal \cite{Sagl97b}).  The diagram is
changed little if instead of $v_c^{\rm max}$ we use the on average
slightly lower circular velocities at $1R_e$ for the elliptical galaxy
points.  Note, however, that there is only a partial overlap in
velocity for the spiral and elliptical galaxy samples.  With the
present data the case for elliptical galaxies having indeed lower
baryonic mass than spiral galaxies of the same circular velocity is
therefore not entirely clear.

It is noteworthy, however, that the baryonic masses of elliptical
galaxies from dynamics are if anything slightly lower
than the baryonic masses of spiral galaxies from realistic stellar population
models, in the region where both distributions overlap. This suggests
that the underlying assumption, that ellipticals are described by
maximum stellar mass models is approximately correct and, hence, that
elliptical galaxy halos have indeed fairly flat cores:  in
hierarchical models in which ellipticals form through merging, a
continuity between spirals and ellipticals would be expected.
Near-maximal M/L in ellipticals are in line with results for the Milky
Way (Gerhard \cite{Gerh99}) and barred galaxies (Debattista \&
Sellwood \cite{DS98}, Weiner, Sellwood \& Williams \cite{WSW00}),
where independent dynamical constraints on the luminous mass favour
near-maximal disks, while the situation is less clear for luminous
unbarred spirals (see, e.g.,. Athanassoula, Bosma \& Papaioannou
\cite{ABP87}, Courteau \& Rix \cite{CouRix99}, Salucci \& Persic
\cite{SP99}, Bell \& de Jong \cite{BdJ00}).

\subsection{Fundamental plane}

In the three-dimensional space defined by central velocity dispersion,
effective surface brightness ($I_e$), and effective radius ($R_e$)
elliptical galaxies fall on a ``fundamental plane'' (FP; Dressler
\etal \cite{Dress+87}, Djorgovski \& Davis \cite{DD87}). The existence
of the FP is thought to be a consequence of the virial theorem
together with a systematic relation of $M/L$ on luminosity (Dressler
\etal \cite{Dress+87}, Faber \etal \cite{Fab+87}). Figure
\ref{fpfitres} shows the FP projection after J{\o}rgensen \etal
(\cite{Joerg96}) for our sample of ellipticals. Effective radii $R_e$
are taken from Table 3 of K+2000, effective surface brightnesses $I_e$
are computed from the (corrected) SB-profiles integrated to $1R_e$
(Table 4 of K+2000), and for the central velocity dispersion we have
used the $\sigma_{0.1}$ defined in \S\ref{secvcsigma}. The
best--fitting line of slope 1.0 is shown along with the data in the
top panel of Fig.~\ref{fpfitres}. The residuals with respect to the
best--fitting line are plotted in the bottom panel against the
velocity anisotropy $\beta_{\rm mean}$.  As Fig.~\ref{fpfitres} shows,
these galaxies follow the FP well and the residuals are not correlated
with $\beta_{\rm mean}$. The rms scatter in $\log R_e$ is 0.084,
excluding NGC 4486B. If we fit the slope as well, it is
$0.924\pm0.069$ and the scatter becomes 0.082.

In the previous section, we have remarked on the correlation of the
residuals from the elliptical galaxy TF relation with $R_e$. This
implies that also in the $v_c^{\rm max}-R_e-B^i_T$ space elliptical
galaxies fall on a fundamental plane. Because $v_c^{\rm max}$ is
very tightly correlated with $\sigma_{0.1}$ (Fig.~\ref{vcsig}),
the resulting plane is very similar to the standard FP.

\subsection{Luminous mass-luminosity relation}
\label{seclml}

\begin{figure*}[t]
\begin{center}
\resizebox{\figwidth}{!}{\includegraphics{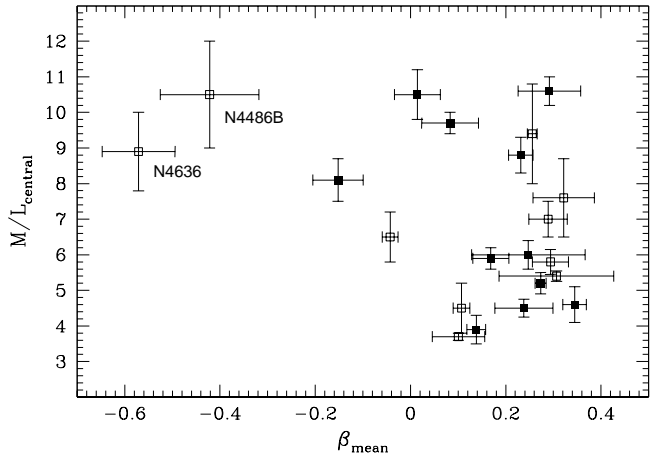}}
\caption[dsb]{\small The central B-band $M/L$ versus mean
velocity anisotropy $\beta$. No correlation is found.}
\label{mlbeta}
\end{center}
\end{figure*}

\begin{figure*}
\begin{center}
\resizebox{\figwidth}{!}{\includegraphics{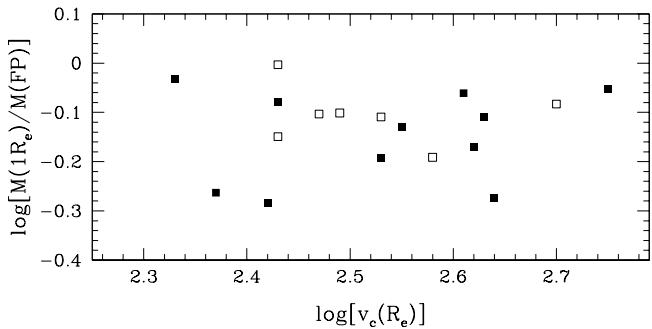}}
\caption[dsb]{\small Ratio of the cumulative mass $M(R_e)$, luminous
and dark, to FP mass $3R_e\sigma_0^2/G$, versus circular velocity at
$R_e$. Symbols as in Fig.~\ref{betmvc}.}
\label{fpratio}
\end{center}
\end{figure*}

\begin{figure*}[t]
\begin{center}
\resizebox{\figwidth}{!}{\includegraphics{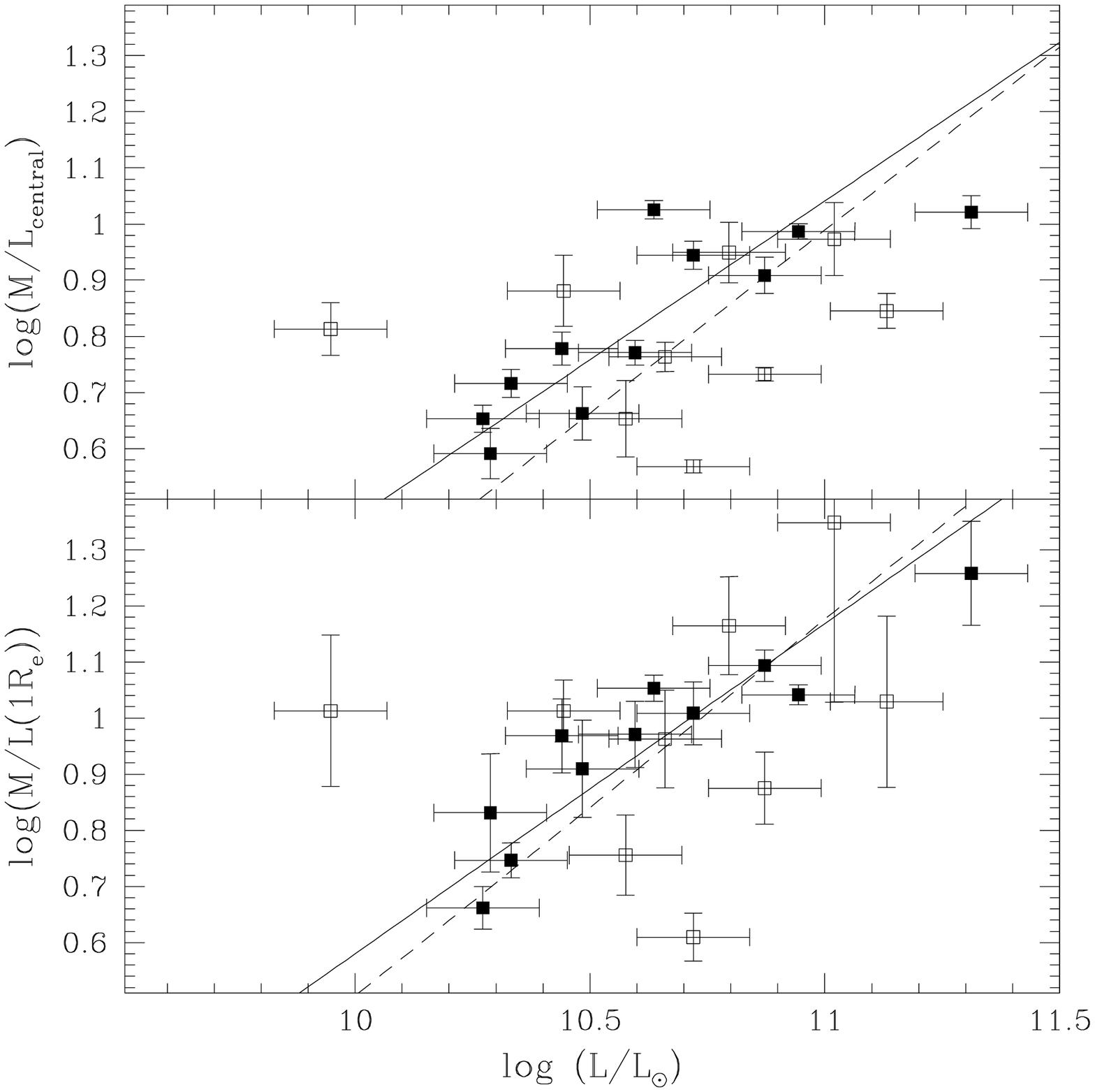}}
\caption[dsb]{\small Mass-to-light ratios as a function of luminosity.
The upper panel shows the log of the central B-band $M/L$ versus total
$L_B$. The lower panel shows the cumulative $M/L_B(R_e)$ at one effective
radius versus $L_B$, including any dark mass. Symbols are as in
Fig.~\ref{betmvc}. Total $L_B$ and errors
as in Fig.~\ref{tfFj}. The errors on the mass-to-light ratios
correspond to one-half the range in the $M/L$-profiles of K+2000,
as determined by dynamical models bounding the respective confidence
interval for each galaxy. In each panel two least-square fits are
shown, for all galaxies in the sample but NGC 4486B (dashed lines),
and for the EK subsample only (full lines). The slopes of these
two fits are not significantly different.}
\label{lmllum}
\end{center}
\end{figure*}

From the variables $(R_e,\sigma_{0.1},I_e)$ one may define a
luminosity $L = c_L I_e R_e^2$ and mass $M_{FP}= 3 c_M
R_e\sigma^2_{0.1}/G$.  As first pointed out by Faber \etal
(\cite{Fab+87}), the virial theorem would then predict $I_e\propto
\sigma^2_{0.1} R_e^{-1}$, provided the mass-to-light ratio and the
structure constants $c_L$ and $c_M$ are identical for all galaxies.
The tilt of the FP with respect to the virial relation therefore
implies either a luminosity dependence of $M/L$ or some deviation of
the family of elliptical galaxies from homology. In the former case
and if the FP is $R_e\propto \sigma^\alpha_{0.1} I_e^\beta$, we would
expect $M/L\propto L^{(2-\alpha)/2\alpha}
I_e^{(-2-\alpha-4\beta)/2\alpha}$.  For the present sample
$\alpha=1.15$ and $\beta=-0.758$, so that one obtains $M/L\propto
L^{0.37} I_e^{-0.05}$.

From our dynamical models we have derived central and cumulative
B-band mass-to-light ratios from observed SB-profiles and kinematics.
The central $M/L_B$ corresponds to the mass-to-light ratio of the
stellar population, on the assumption of a maximum stellar mass
(minimum halo) model in which the luminous stars provide as much mass
in the central parts as is allowed by the kinematic data.  These
$M/L_B$ values have no residual correlation with anisotropy
(Fig.~\ref{mlbeta}); this is expected as the anisotropy was taken into
account in the modelling.

Figure~\ref{fpratio} shows the ratio of the dynamically determined
cumulative mass at $1R_e$, including any dark mass that may be
necessary up to this radius, and the FP mass, $M_{FP}\equiv
3R_e\sigma_0^2/G$, for galaxies with data to beyond $R_e$. This ratio
measures the structure constant $c_M$. From the figure it is clear
that there is no systematic trend of $c_M$ with circular velocity or
mass. Because both photometry and dynamical anisotropy were taken into
account in the modelling, this shows that the dynamical non-homology
mechanism proposed by Graham \& Colless (\cite{GC97}) is not the main
cause for the tilt of the FP.  On the contrary, the dynamical structure of
elliptical galaxies is remarkably uniform (Fig.~\ref{vcsig}), and 
no correlation was found between $M/L$ and anisotropy (Fig.~\ref{mlbeta}).

Figure~\ref{lmllum} shows the dynamically determined mass-to-light
ratios against luminosity. The upper panel of Fig.~\ref{lmllum} shows
the central $M/L_B$, i.e. the inferred mass-to-light ratio of the stellar
population in our models. The range in central $M/L_B$ is about a
factor of three. The lower panel shows the cumulative
mass-to-light ratio at $R_e$. The scatter in these diagrams is
comparable to eachother, and galaxies with particularly low or high
$M/L_B$ correspond almost one by one. For the EK sample the scatter is
similar to the scatter in the FP mass-luminosity relation while for
the BSG sample it is somewhat larger, as might be expected.

Contrary to the range in central $M/L_B$, which is fixed by the
dynamical models, the actual slope of the $M/L-L$ relation also
depends on the extrapolation used to derive the total luminosity; the
K+2000 values were derived from the photometry using a de Vaucouleurs
law. The fitted slopes are slightly steeper than the predicted FP
relation. For the EK sample they are for $M/L_{\rm central}$: $0.57\pm
0.11$ when the errors are rescaled to $\chi^2=1.0$, and for
$M/L(R_e)$: $0.59\pm0.09$ with $\chi^2=0.7$. For the total sample
(excluding NGC 4486B) they are for $M/L_{\rm central}$: $0.65\pm 0.09$
when $\chi^2=1.0$, and for $M/L(R_e)$: $0.67\pm0.15$ when
$\chi^2=1.0$.  These slopes are different from the predicted FP slope
by 1.5-2.5 times the $\sigma$ of the $M/L$ fit, and they are either
influenced by small number effects (for the EK sample) or by outlying
data points (for the full sample). They thus appear still consistent
with the predicted FP slope.  Fig.~\ref{vcsig} and Fig.~\ref{fpratio}
which demonstrate the dynamical similarity of our galaxies also
suggest that the difference may not in fact be significant.

From these results we conclude that the $M/L$ variations with
luminosity indicated by the tilt in the fundamental plane are real.
The inferred FP $M/L$ ratios for our sample of ellipticals 
correspond to the dynamically measured
$M/L$ values for the stellar populations, assuming a minimal halo, and
they are not due to a non-homologous dynamical structure changing
gradually with luminosity. It is still possible, however, that photometric
non-homology influences the slope of the $M/L-L$-relation.  The trend
of $M/L$ with $L$ is also not caused by an increasing fraction of dark
matter as luminosity increases, unless two thirds of the luminous mass
in the cores of the most massive galaxies is dark matter, and luminous
elliptical galaxies are then significantly short of baryons compared
to spirals (see \S\ref{sectfbary} and Fig.~\ref{tfbary}). The change
of $M/L$ with $L$ is therefore most likely due to the population
itself.


\section{Central $M/L$ and stellar populations}
\label{secpop}

\begin{table*}[t]
\caption[M/L]{The $M/L_B$ values of the stellar populations in solar
units, using the models of Maraston (\cite{M98}).  Col. 1 gives the
galaxy name, Col. 2 the $M/L_B$ of the luminous component from our
dynamical analysis, using $H_0=65\kms\mpc^{-1}$, Col. 3 the $M/L_B$
derived using the metallicities of Kobayashi \& Arimoto (\cite{KA99},
KA) with 15 Gyr age, Col. 4 using ages and metallicities from
Terlevich \& Forbes (\cite{TF00}, TF), and Col. 5 using ages and
metallicities inside $(R_e/8)$ of Trager et al. (\cite{Trager+00a},
\cite{Trager+00b}, TFWG). Columns 3, 4 and 5 are computed for a
Salpeter IMF. Cols. 6, 7 and 8 repeat Cols. 3, 4 and 5, but for the
Kroupa (\cite{Kr00}) IMF. Cols. 9, 10 and 11 are the same for the GBF IMF.}
\begin{flushleft}
\begin{tabular}{rrrrrrrrrrrrr}
\noalign{\smallskip}
\hline
\noalign{\smallskip}
 & & \multicolumn{3}{c}{Salpeter} & \multicolumn{3}{c}{Kroupa} &
\multicolumn{3}{c}{GBF}\\
Galaxy   & Dyn.  &  KA     &    TF  & TFWG    &  KA     &    TF  & TFWG    &  KA     &    TF  & TFWG      \\
          &      & (15 Gyr)&        & $(R_e/8)$ & (15 Gyr)&        & $(R_e/8)$ & (15 Gyr)&        & $(R_e/8)$ \\
\noalign{\smallskip}
\hline
\noalign{\smallskip}
NGC  315 & 10.5  &10.1 &    5.8 & 6.5   & 6.4 & 3.3 & 3.8 & 5.0 & 3.2 & 3.5\\
NGC 1399 & 10.6  & -   &    5.8 & 12.6  & -   & 3.3 & 7.5 & -   & 3.2 & 6.8\\
NGC 2434 &  6.0  & 5.8 &    -   &    -  & 3.8 & -   & -   & 3.0 & -   & -  \\
NGC 3193 &  4.5  & -   &    5.1 &   -   & -   & 3.2 & -   & -   & 2.5 & -  \\
NGC 3379 &  4.5  & 8.4 &    8.6 & 8.8   & 5.4 & 5.1 & 5.3 & 4.2 & 4.5 & 4.6\\
NGC 3640 &  3.7  & -   &    4.4 &   -   & -   & 2.8 & -   & -   & 2.2 & -  \\
NGC 4168 &  5.8  & -   &    5.2 &   -   & -   & 3.3 & -   & -   & 2.6 & -  \\
NGC 4278 &  7.6  & 8.5 &    6.9 &   -   & 5.5 & 4.3 & -   & 4.3 & 3.5 & -  \\
NGC 4374 &  8.8  & 9.3 &    9.1 & 11.8  & 6.0 & 5.7 & 7.4 & 4.7 & 4.6 & 6.0\\
NGC 4472 &  7.0  &10.9 &    8.5 &  8.3  & 6.9 & 5.1 & 5.0 & 5.4 & 4.5 & 4.4\\
NGC 4486 &  9.4  &11.1 &     -  &    -  & 7.0 & -   & -   & 5.5 & -   & -  \\
NGC 4494 &  6.5  & 5.8 &     -  &    -  & 3.8 & -   & -   & 3.0 & -   & -  \\
NGC 4636 &  8.9  & 8.4 &     -  &   -   & 5.4 & -   & -   & 4.2 & -   & -  \\
NGC 5846 &  9.7  &12.1 &    11.9&  13.5 & 7.7 & 7.2 & 8.5 & 6.0 & 6.2 & 6.8\\
NGC 6703 &  5.2  & -   &     4.8&  4.9  & -   & 2.7 & 2.8 & -   & 2.7 & 2.7\\
NGC 7192 &  4.6  & 6.3 &     -  &   -   & 4.1 & -   & -   & 3.2 & -   & -  \\
NGC 7626 &  8.1  & 9.6 &    10.6&  13.0 & 6.2 & 6.6 & 8.1 & 4.8 & 5.4 & 6.6\\
\noalign{\smallskip}
\hline
\noalign{\smallskip}
\end{tabular}
\end{flushleft}
\label{tabml}
\end{table*}

\begin{figure*}
\begin{center}
\resizebox{\figwidth}{!}{\includegraphics{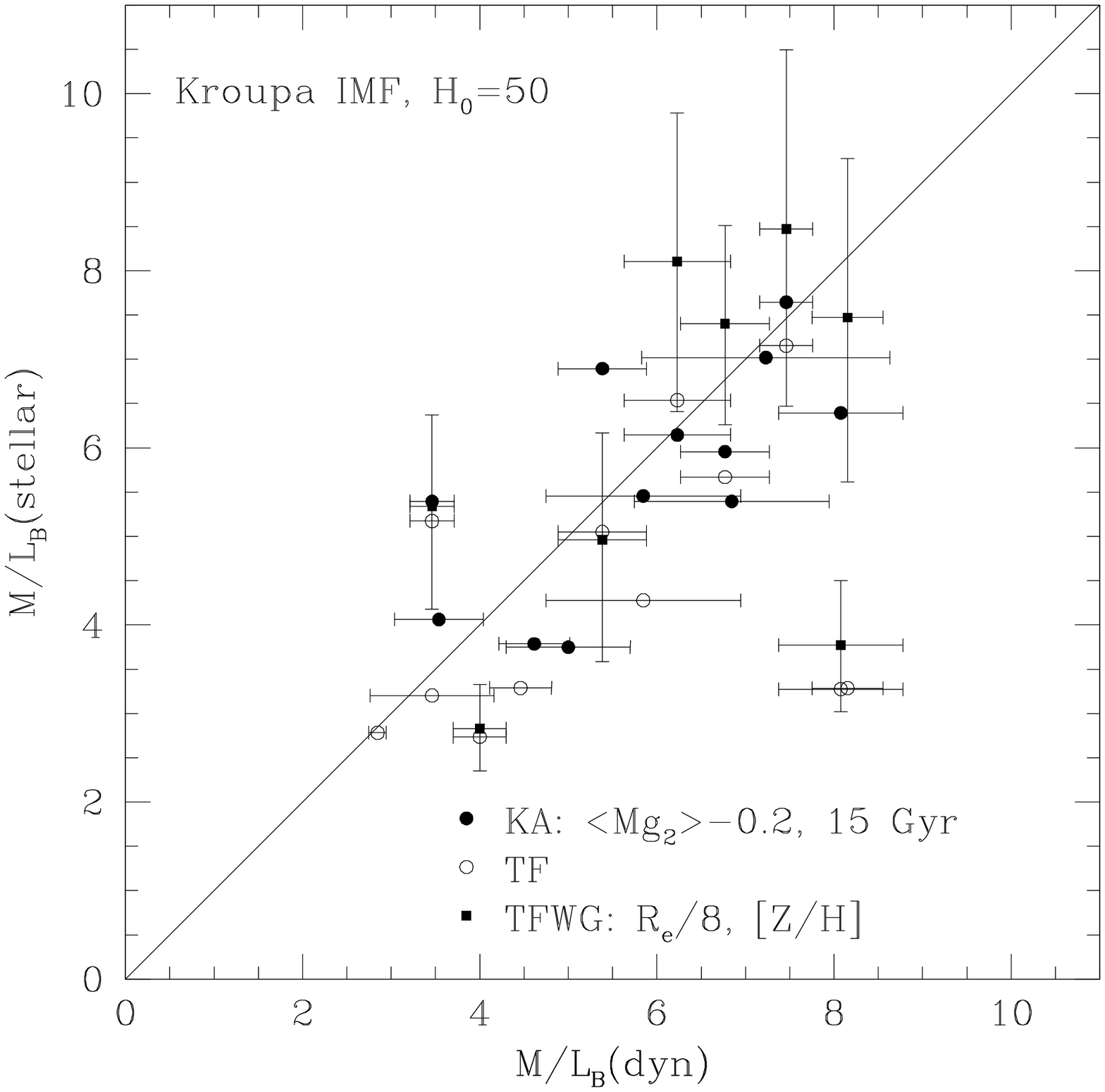}}
\caption[mlmod]{\small The comparison between the dynamical estimates
of the central B-band $M/L$ (transformed to $H_0=50\kms\mpc^{-1}$) with the
predictions of stellar population models of Maraston (\cite{M98}),
using Kroupa's (\cite{Kr00}) IMF. 
Metallicities and ages used in the models are 
taken from Kobayashi \& Arimoto (\cite{KA99},
filled circles),  Terlevich \& Forbes (\cite{TF00}, open circles), 
Trager et al. (\cite{Trager+00a}, \cite{Trager+00b}, filled squares). }
\label{figmlmodels}
\end{center}
\end{figure*}

The question of what causes the variation with luminosity of the
$M/L$s observed in ellipticals has been discussed at length in the
literature, without any fully satisfactory answer. Systematic
deviations from homology have been argued to play a role (Pahre et
al. \cite{Petal98}), but as discussed above this is not a viable
solution. It is possible to fit the trend observed in the B-band as a
metallicity sequence of an old stellar population (Maraston
\cite{M99}). However, the $M/L$s in the K band which this model would
predict are independent of metallicity, and therefore no correlation
with K-band luminosity should be expected, contrary to what is
observed (Pahre et al. \cite{Petal98}). Forbes et al. (\cite{Fetal98})
and Forbes \& Ponman (\cite{FP99}) analyse the ages determined for 88
galaxies and conclude that the observed correlation between age and
luminosity is far too weak to explain the observed $M/L$ trend with
luminosity in any band.

Here we explore whether stellar population models can reproduce the
central B-band $M/L_B$ values derived from our (minimum halo)
dynamical models.  We compare these $M/L_B$ values with the
predictions of stellar populations models in Fig. \ref{figmlmodels}.
We use the stellar population models of Maraston (\cite{M98}), as
employed in Saglia et al.\ (\cite{Setal00}), to interpolate the
$M/L_B$ expected for a simple stellar population of given metallicity
and age.  These models take into account stellar evolution mass loss,
i.e., they include in the mass budget the masses of stellar remnants,
but not the mass losses of the progenitor stars.  This is different
from what is done by, e.g., Worthey (\cite{W94}) or Bruzual \& Charlot
(\cite{BC96}), where the total initial mass is conserved.  In addition
to the classical Salpeter IMF (with power law exponent $\gamma=-2.35$
independent of stellar mass), we consider two additional choices for
the IMF: the
recent comprehensive determination of Kroupa (\cite{Kr00}), indicating
that the low mass stars could be less numerous (his eq.~(2):
$\gamma=-1.3$ for $m<0.5 m_\odot$, $\gamma=-2.3$ at larger masses),
and the more extreme IMF of Gould, Bahcall and Flynn (\cite{GBF97},
hereafter GBF), where a flatter slope at lower masses is suggested
(after correction for binaries, $\gamma=-0.9$ for $m<0.6 M_\odot$,
$\gamma=-2.21$ for $0.6<m<1$ and $\gamma=-2.35$ for $m>1 M_\odot$).
For all three IMFs we use a lower stellar mass cut off of $0.1
M_\odot$.

As input we use ages and metallicities as derived by Kobayashi \&
Arimoto (\cite{KA99}), Terlevich \& Forbes (\cite{TF00}) and Trager et
al. (\cite{Trager+00a}). Such data are available from at least one
study for 17 out of 21 of our galaxies.

Kobayashi \& Arimoto (\cite{KA99}) derived mean metallicities inside
$1R_e$, considering line index gradients. We use values as given in
their Table 2. These are derived from the $Mg_2$ line, assuming an age
of 17 Gyr. The $M/L_B$ values shown in Fig. \ref{figmlmodels} are
determined by reducing their metallicities by 0.2 dex (to correct for
the Mg/Fe overabundance) and for an age of 15 Gyr.

Terlevich \& Forbes (\cite{TF00}) compiled a catalogue of high quality
absorption line measurements for galaxies and derived separate age
and metallicity estimates using Worthey (\cite{W94}) models.

Trager et al. (\cite{Trager+00a}) determined ages and metallicities
after applying a correction for the Mg/Fe overabundance to the line
indices. We use their ages and metallicities for the inner $R_e/8$
datapoints, and average over the 4 models considered by these authors,
for all galaxies except NGC 1399. For this galaxy we use the new age
and metallicity determination by Trager et al. (\cite{Trager+00b});
for the other galaxies the new determinations agree with the previous
values within the errors.

Table \ref{tabml} lists the resulting stellar population $M/L_B$
values with each of the three IMFs. Fig. \ref{figmlmodels} shows the
comparison of the stellar M/Ls determined with Kroupa's (\cite{Kr00})
IMF with the dynamical $M/L_B$s, rescaled to $H_0=50\kms\mpc^{-1}$ which
gives the best overall agreement.  The galaxy-by-galaxy comparison
after this rescaling is reasonable within the errors.  Only two
objects are particularly deviant, NGC 315, where the dynamical value
is a factor 2 larger than the stellar population estimates based on
Terlevich \& Forbes (\cite{TF00}) and Trager et al.
(\cite{Trager+00a}), but is in agreement with the value obtained from
Kobayashi \& Arimoto (\cite{KA99}), and NGC 1399, where $M/L_B$(dyn)
is again a factor 2 larger than the stellar $M/L_B$ based on Terlevich
\& Forbes (\cite{TF00}), but agrees with the value derived from Trager
\etal (\cite{Trager+00b}).  This discrepancy is due to the rather low
ages (5 Gyr) assigned there, which is also in conflict with the
determinations of Maraston and Thomas (\cite{MT00}).

Similar plots are obtained for the Salpeter and the GBF IMFs, when
distances are scaled to an optimal $H_0=75\kms\mpc^{-1}$ for the former
and $H_0=40\kms\mpc^{-1}$ for the latter case.  These diagrams show that
the dynamical $M/L_B$ obtained with $H_0=65\kms\mpc^{-1}$ are $\approx
30$\% larger than the stellar M/Ls with the Kroupa (\cite{Kr00}) IMF,
$\approx 60$\% larger with the GBF IMF, and $\approx 20$\% smaller
than the stellar M/Ls with the Salpeter IMF.  Note that an IMF flatter
than Salpeter for $m>1M_\odot$ produces stellar M/Ls larger than the
Salpeter values.

We conclude that our dynamical $M/L_B$ values, based on models
maximizing the contribution of the luminous component, are compatible
with those predicted by stellar population models, within the
uncertainties in the distance scale and the poorly known fraction
of low-mass stars present in giant ellipticals.  Only in the case that
(i) an IMF as flat at low stellar masses as that of GBF is applicable
to our elliptical galaxies, and simultaneously (ii) a short distance
scale ($H_0\simeq 80\kms\mpc^{-1}$) turns out to be correct, would we have
underestimated the luminous masses by as much as a factor $\approx 2$
by making the assumption of maximum stellar, or minimum dark halo
mass. For lower values of $H_0$ and/or the other IMFs investigated the
difference is smaller. For comparison, recall that to explain the tilt
of the FP as due to an increasing fraction of dark matter with
luminosity, we would require the stellar mass in the most luminous
ellipticals to be only one third of the inferred dynamical mass.
Furthermore, the ratio of $M/L_B$(dyn)$/ M/L_B$(stellar) for this
sample of elliptical galaxies does not correlate with luminosity.

Together with the results of Section \ref{secscalings} this suggests
that the tilt of the FP is a stellar population effect. Most likely
the main driver is metallicity, but then a secondary population effect
is needed to explain the K-band tilt (Pahre \etal \cite{Petal98}). A
larger sample of galaxies with detailed dynamical modeling also of the
K-band profiles will be needed to decide whether this is the final
answer to the problem of the tilt of the Fundamental Plane.


\section{Mass-to-light ratios and minimum halo properties}

\begin{figure*}[t]
\begin{center}
\resizebox{\figwidth}{!}{\includegraphics{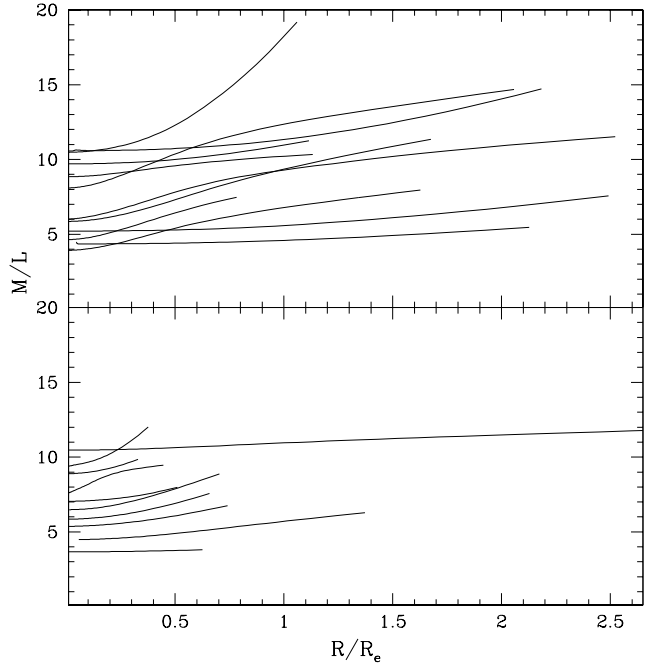}}
\caption[dsb]{\small Cumulative B-band $M(r)/L_B(r)$ as function of
radius. Upper panel: EK galaxies, lower panel: BSG galaxies.}
\label{mlRAll}
\end{center}
\end{figure*}

\begin{figure*}[t]
\begin{center}
\resizebox{\figwidth}{!}{\includegraphics{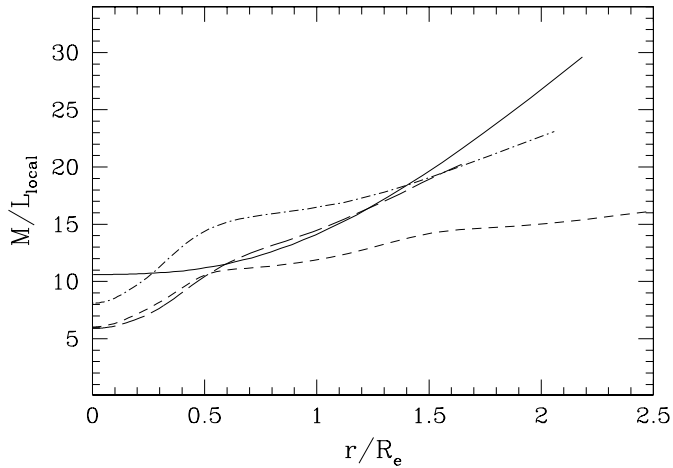}}
\caption[dsb]{\small {\sl Local} B-band $M/L_B$ as function of radius
for the four most reliable galaxies from the EK sample: NGC 1399
(full line), NGC 7626 (dot-dashed), NGC 7507 (long-dashed), and
NGC 2434 (short-dashed), over the range of the modelled 
kinematic data.
}
\label{mlRlocal}
\end{center}
\end{figure*}

In this section we first show that the dynamical models imply
significant radial $M/L$ variations in elliptical galaxies. Because
the implied {\sl local} mass-to-light ratios are large, we argue that
these are the signature of dark matter halos.  Based on the result of
the previous section, that the dynamically determined central $M/L_B$
values for our ellipticals are consistent with the stellar population
$M/L_B$s, we use the dynamical models to delineate the decomposition
of the elliptical galaxy circular velocity curves (CVCs) into luminous
and dark contributions, and finally study the properties of the
implied minimum halos.

\subsection{Global and local mass-to-light ratios}

Figure \ref{mlRAll} shows the radial profiles of cumulative
$M(r)/L_B(r)$ for all galaxies in the two subsamples, as derived from
the respective ``best model'' CVC (Fig.~\ref{vcall}) and luminosity
profile. There is considerable variety in these cumulative
$M/L_B$-profiles; even within the galaxies with extended kinematics,
the ratio of cumulative $M/L_B$ at the outer data boundary to the central
value ranges from consistent with 1 to $\lta 2$ (see also Table 7
of K+2000). This suggests a corresponding spread in the efficiency
of dissipational segregation and angular momentum loss of the
stellar and gaseous component in the dark matter halo during the
formation process.

Even though the gradients in cumulative $M/L_B$ are modest, those in
the {\sl local} mass-to-light ratio $\rho(r)/j_B(r)$ are not. (Here
$\rho(r)$ and $j_B(r)$ are the inferred mass and luminosity
densities). Because these local $\rho/j$-ratios are less certain than
the cumulative $M/L_B$-values we show these in Figure \ref{mlRlocal}
only for the best cases from the EK-subsample with extended
kinematics.  As the figure shows, these local $M/L_B$s become large
$\sim 20-30$ in the modelled outer parts of these galaxies. This, and
corroborating evidence from X-ray data (e.g., Matsushita \etal \cite{Metal98},
Loewenstein \& White \cite{LW99}) argues strongly that the measured $M/L$
variations are not due to a slow outward change of the stellar
population, but instead imply dark matter halos similar to those
inferred in spiral galaxies, where the component contributing most of
the mass, baryonic or not, is very different from a normal stellar
population.

\begin{figure*}
\begin{center}
\resizebox{\figwidth}{!}{\includegraphics{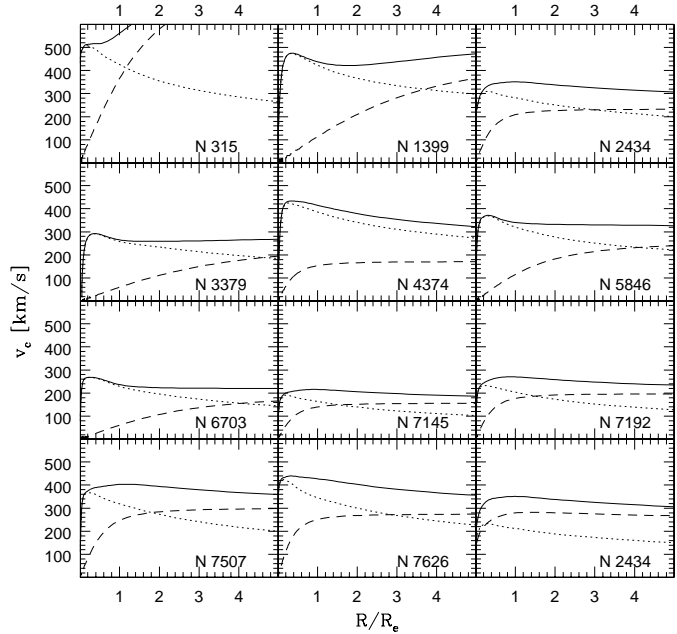}}
\caption[dsb]{\small Total circular velocity curves (solid lines),
and contributions
of the luminous matter (dotted line) and the dark halo (dashed). Note
that the curves are extrapolated to $5 R_e$.}
\label{vcCurves1}
\end{center}
\end{figure*}

\subsection{Circular velocity curve decomposition}

Thus, as for spiral galaxies, it is of interest to analyze the
relative contributions of the luminous and dark matter components to
the CVCs in Fig.~\ref{vcall}. In doing this we assume that the $M/L_B$
of the luminous component is constant with radius and has the maximal
value allowed by the kinematic data, providing nearly all the mass in
the center.  The dynamical models used by K+2000 to analyze the
kinematic data were built on this assumption, but once the CVCs are
determined, the luminous component could in principle be assigned less
mass a posteriori.  The discussion in \S4 has shown that the maximum
central $M/L_B$ values determined for our ellipticals by K+2000 are
consistent with the $M/L_B$ values expected for the stellar population
of these galaxies, within the uncertainties in the distance scale and
the lower-mass IMF. Fig.~\ref{tfbary} has also shown that even with
maximal $M/L_B$ elliptical galaxies have if anything slightly lower
baryonic mass than spiral galaxies of the same circular velocity.

Figure \ref{vcCurves1} shows the total circular velocity curves from
Fig.~\ref{vcall} for all galaxies in the high-quality EK-sample, and
the respective contributions from the maximum luminous and
corresponding minimum dark halo components. At $1R_e$, the halo
contributes between $1/4$ to $2/3$ of the circular velocity in these
``best'' dynamical models, corresponding to between $10-40\%$ of the
integrated mass.  All curves are plotted to radii of $5R_e$,
extrapolating the models beyond the radial range of the data. The case
of NGC 315 (where the outer rise of the CVC is due to modelling
problems, see K+2000) shows that such extrapolation can lead to large
errors. However, in cases where X-ray data or planetary or globular
cluster velocities were available, the ``best'' models of K+2000
matched the independent mass estimates from these outer data very
well. Figure \ref{vcCurves1} shows that within the framework of the
models, luminous and dark matter reach equal interior mass at $\sim
2-4 R_e$, and at $5R_e$ the halo is predicted to dominate in all
models except in one case. As in spiral galaxies, the combined
rotation curve is flatter than that for the individual components
(``conspiracy''); this is already seen within the radial range of the
kinematic data.

The last panel of Figure \ref{vcCurves1} repeats the CVC decomposition
for NGC 2434 with a luminous $M/L_B=3.4$ instead of the maximum
$M/L_B=6.0$. This is the average of the population values obtained for
this galaxy with the Kroupa and GBF IMF's (see Table \ref{tabml}).
The lower $M/L_B$ leads to a significantly denser halo; the
decomposition in this case is comparable to those of Rix \etal
\cite{Retal97}) with a Navarro, Frenk \& White (NFW, \cite{NFW96})
halo mass distribution. This example suggests (i) that the kinematic
data for the K+2000 sample could presumably have been fit also by
using these halo models, and (ii) that the resulting lower $M/L_B$
values for the luminous components would be near the lower end of the
range consistent with the stellar population models of Section
\ref{secpop}. However, in this case, ellipticals would move further down
in Fig.~\ref{tfbary}: in the models of Rix \etal
(\cite{Retal97}) with a NFW halo for NGC 2434 the $M/L$ for the
luminous component is another (for the same distance) factor of 1.5
smaller than the maximum $M/L$ found by K+2000, which had already
placed NGC 2434 by a factor of 4 below the Bell \& de Jong 
(\cite{BdJ00}) and McGaugh \etal (\cite{McG00}) lines in
Fig.~\ref{tfbary} (at $\log v_c^{\rm max}=2.55$).

In a recent study, Loewenstein \& White (\cite{LW99}) considered the
implications of the observed X-ray temperature - velocity dispersion
relation for the mass distribution and dark halos in elliptical
galaxies. They found that the total (luminous plus dark) $M/L_V$ at
$6R_e$ is nearly independent of galaxy luminosity, with value
$M/L_V(6R_e)\simeq 25 h_{80} \msun/L_{\odot, V}$, so that the ratio of
dark to luminous matter decreases with luminosity. Their $M/L(6R_e)$
converted to the B-band and rescaled to $h_{65}$ is consistent with
the $M/L_B(6R_e)$ predicted by our models for $L_B\gta L_\ast$
galaxies, while for our fainter galaxies the dynamical models predict
$\sim 0.2$ dex lower mass per luminosity than given by Loewenstein \&
White (\cite{LW99}), in all cases, however, extrapolating the models
beyond the radial range of the kinematic data. The dark matter mass
fraction within $R_e$ found from the X-ray analysis ($\gta 20\%$) is
in good agreement with our result quoted above.

\subsection{Dark halo parameters for ellipticals}

\begin{figure*}[t]
\begin{center}
\resizebox{\figwidth}{!}{\includegraphics{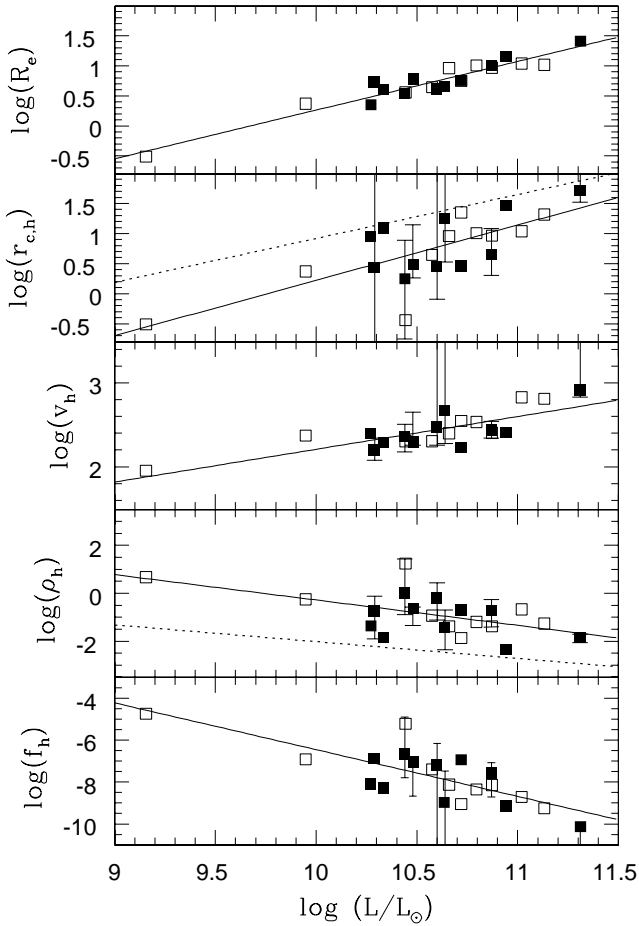}}
\caption[dsb]{\small From top to bottom: (i) Effective
radius in kpc versus total luminosity in solar
units. (ii) Halo core radius in kpc, (iii) halo velocity in $\kms$,
(iv) central halo density in $\msun/\pc^3$, and (v) central halo
phase-space density in $\msun/\pc^3/(\kms)^3$, all referring to the
``best'' models selected by K+2000 from the middle of
the respective confidence range for all sample galaxies. Symbols are
as in Fig.~\ref{betmvc}. The full lines show the
least-square fits. Dotted lines show the corresponding relations for
spiral galaxies given in the text.
}
\label{rcLum}
\end{center}
\end{figure*}

In this section we investigate scaling laws for the halo densities and
halo core radii of elliptical galaxies based on the
``minimum halo models'' of K+2000. We also compare the inferred halo
properties with the dark matter halos in spiral galaxies.

Figure \ref{rcLum} shows, as a function of galaxy luminosity, the
effective radius $R_e$, the halo core radius $r_{c,h}$, the halo
velocity $v_h$, core density $\rho_h$, and central phase-space density
$f_h$, for the respective ``best'' models of K+2000. All these
correlate with luminosity, but the tightest correlation is that
between the effective radius and the luminosity (one of the
projections of the FP).

We note that for many of the galaxies in Fig.~\ref{rcLum} a constant
$M/L_B$ model with only luminous mass is within the 95\% confidence
range of the K+2000 models. For these galaxies the halo radii, circular
velocities, and densities have therefore large uncertainties in a
logarithmic plot like Figure \ref{rcLum}. However, in cases where
X-ray data or planetary or globular cluster velocities were available,
the ``best'' models of K+2000 matched the independent mass estimates
from these outer data very well, while the constant $M/L$ models were
usually inconsistent with these data.  For seven of the sample
galaxies, a constant $M/L$ model was found inconsistent with the
kinematic data
(K+2000). For these galaxies we have estimated $95\%$ confidence
bounds on the halo parameters from $\chi^2$ contour plots and plotted
them as error bars in Fig.~\ref{rcLum}.  In a few of these cases only
the halo density is well-determined.  Within the considerable
uncertainties, the `error bars' are consistent with the scatter of the
points.  Note that some of the galaxies with the best evidence for
dark halos (NGC 2434, 7507, 7626) are among the smallest halo core
radii and largest halo density points with respect to the mean fit
lines in Fig.~\ref{rcLum}. By using the mean scaling relations from
all the ``best'' halo models of K+2000 for the subsequent discussion,
we have therefore not biassed the normalisations of the best-fit
lines, while the best-fit slopes are much better determined.

Figure \ref{rcLum} shows that, in the mean,
more luminous galaxies have larger halo core radii, with a slope
similar to the $\log(R_e)$-$\log(L)$-relation. They also have larger
halo circular velocities, and lower central densities and phase-space
densities. As predicted by hierarchical models less massive objects
are denser. The least-square fits shown in Figure \ref{rcLum}
correspond to the following scaling laws:
\begin{equation}\label{eqReL}
 R_e = 11.8 L_{11}^{0.81} h_{0.65}^{-1} \kpc
\end{equation}

\begin{equation}\label{eqrhL}
 r_{c,h} = 13.8 L_{11}^{0.92} h_{0.65}^{-1} \kpc
\end{equation}

\begin{equation}\label{eqvhL}
 v_h = 397 L_{11}^{0.39} \kms
\end{equation}

\begin{equation}
 \rho_h = 0.046 L_{11}^{-1.06} h_{0.65}^{2} \msun/\pc^3
\end{equation}

\begin{equation}\label{eqfhL}
 f_h = 2.0 \times 10^{-9} L_{11}^{-2.23} h_{0.65}^{2} \msun/\pc^3/(\kms)^3
\end{equation}
where $L_{11}=L_B/10^{11}h_{0.65}^{-2} L_{\odot,B}$, 
\begin{equation}\label{eqrhoh}
 \rho_h = {3\over 4\pi G} {v_h^2\over r_{c,h}^2}
\end{equation}
for the employed halo models (see eqs.~(2-4) in Gerhard \etal \cite{G+98}), and
the central phase-space density is defined by
\begin{equation}\label{eqfh}
 f_h \equiv 2^{3/2} \rho_h / v_h^3.
\end{equation}
Eqs.~(\ref{eqReL}) and (\ref{eqrhL}) result in a ratio of $R_e$
and halo core radius that is approximately constant,
\begin{equation}\label{eqRerhL}
 r_{c,h}/R_e = 1.2 L_{11}^{0.11}.
\end{equation}
Also, from the results of \S\ref{seclml} and eqs.~(\ref{eqReL}) 
and (\ref{eqrhoh}) one sees that
the ratio of luminous mass density $(M/L_B) L_B/R_e^3$ and dark
halo density has little luminosity dependence in the mean, although
the scatter is large; the density ratio varies between $10^{\sim(0.5-2)}$.

The dotted lines in Fig.~\ref{rcLum} show the scaling relations for
spiral galaxies from Persic, Salucci \& Stel (\cite{PSS96a}, erratum
\cite{PSS96b}; PSS), which are based on the same (minimum) halo
models.  The relation for spiral galaxy halo central densities as
given by PSS and rescaled to the distance scale used here becomes
\begin{equation}\label{eqrhohS}
 \rho^S_h = 0.0019 L_{11}^{-0.7} h_{0.65}^{2} \msun/\pc^3.
\end{equation}
For luminosities around $L_B\simeq 10^{11}L_{B,\odot}\simeq 3L_\ast$,
elliptical galaxy halos are therefore about 25 times denser than
spiral galaxy halos of the same $L_B$, assuming maximum stellar
mass in both cases. This result agrees well with
the work of Bertola \etal (\cite{BPPS93}), who used extended HI disks
around a few elliptical galaxies to constrain their halo mass
distributions.  Most of the factor 25 can be traced back to the fact
that the CVCs of both spirals and ellipticals are approximately flat,
and that for an elliptical galaxy profile, the maximum circular
velocity occurs at significantly smaller radius in units of $R_e$ than
for an exponential disk. There is an additional factor $\sim 2$
because of the larger $M/L_B$ of ellipticals at given $L_B$.

According to PSS, the halo core radii of spiral galaxies also scale
with luminosity when expressed in units of the optical radius. We can
compare their relation to the case of elliptical galaxies as follows:
Fitting the parameters $R_e$ and $L$ for $\sim 200$ spiral galaxies
from the RC3, as given by Burstein \etal (\cite{BBFN97}), results in the
least-square fit line (rescaled to $H_0=65 \kms\mpc^{-1}$)
\begin{equation}
  R_e^S = 9.0 L_{11}^{0.53} h_{65}^{-1} \kpc.
\end{equation}
This is somewhat shallower than the corresponding relation for
elliptical galaxies, eq.~(\ref{eqReL}), but the $R_e$ values of spirals
and ellipticals are very similar around $L=L_\ast$. Next, we use
$R^S_{\rm opt}\equiv 3.2 R_D=1.9 R^S_{1/2}$, where $R^S_{\rm opt}$,
$R_D$ and $R^S_{1/2}$ are the optical, scale and half-mass radius for
an exponential disk, and assume a mean $R_e^S=1.2 R_D$ for the
distribution between face-on and edge-on.
From the relation given by PSS we then obtain
\begin{equation}\label{eqRerhS}
 r^S_{c,h}/R^S_e = 5.0 L_{11}^{0.2}
\end{equation}
and hence
\begin{equation}\label{eqrhS}
 r^S_{c,h} = 45 L_{11}^{0.73} h_{0.65}^{-1} \kpc.
\end{equation}
Comparing these relations with eqs.~(\ref{eqRerhL}) and (\ref{eqrhL}),
one sees that the minimum halos of spiral galaxies of the same $L_B$
and $R_e$ have about 4 times larger halo core radii than are inferred
for elliptical galaxies from our dynamical analysis, and that the
ratio is only slightly smaller at given $L_B$ if the mean $R_e^S$
[eq.~(\ref{eqrhS})] is used.  In view of the uncertainties in the
transformations, the slopes in these relations appear consistent with
eachother.

Because of the luminosity offset in the TF relation, it may be more
appropriate to compare elliptical and spiral galaxy halos at the same
baryonic mass or circular velocity than at the same luminosity.
Because this would mean comparing an elliptical galaxy with a spiral
of {\sl higher} luminosity, and because of the luminosity dependences
in the spiral galaxy relations eqs.~(\ref{eqrhS}) and (\ref{eqrhohS}),
this will {\sl increase} the differences found above. Using $v_c^{\rm
  max}$ for the comparison as in Fig.~\ref{tfFj}, the density ratio
$\rho_h/\rho_h^S$ thus increases by a factor of $\sim 2$.  This may be
an overestimate, however, as the inferred asymptotic halo velocities
$v_h$ are formally lower than $v_c^{\rm max}$ [by $\sim 0.1$ dex,
compare eqs.~(\ref{eqtfnorm}) and (\ref{eqvhL})]. We note here that
the $v_h$ value for $L_B^\ast$ predicted by eq.~(\ref{eqvhL}),
$253\kms$, is in agreement with the galaxy-galaxy lensing result of
Wilson \etal (\cite{WKLC00}) at radii $\sim 100\kpc$, but emphasize
that in our models $v_h$ is much more uncertain than $v_c^{\rm max}$.
Making the comparison at constant baryonic mass would increase the
density ratio by a factor less than 1.6, from Fig.~\ref{tfbary}
and eq.~\ref{eqrhohS}. Thus we conclude conservatively that the halos
of elliptical galaxies are {\sl at least} 25 times denser than the
halos of spiral galaxies of similar baryonic mass or circular
velocity.

These results also suggest that the phase-space densities for the
minimum halos of elliptical galaxies are higher than for those of
spiral galaxies of similar baryonic mass.  From the TF relation spiral
galaxies at given $L_B$ have circular velocities about 0.2 dex lower
than the $v_c^{\rm max}$ of elliptical galaxies, but perhaps only 0.1
dex lower than the $v_h$ values used in eqs.~(\ref{eqvhL}) and
(\ref{eqfhL}). This is not sufficient to compensate their higher
densities. Higher phase-space
densities would rule out stronger adiabatic contraction as the
explanation for the denser halos in ellipticals. Recall also the
similar $R_e$ of ellipticals and spirals of the same $L_B$.  The
argument could be circumvented if spiral galaxies had sub-maximal
disks and cuspy halos and their halo densities had been underestimated
significantly by the PSS models. However, note that both in the Milky
Way (Gerhard \cite{Gerh99}) and in barred galaxies (Debattista \&
Sellwood \cite{DS98}, Weiner \etal \cite{WSW00}), where independent
dynamical constraints on the luminous mass are available, the galactic
disks are near-maximal, and that the recent stellar population models
of Bell \& de Jong (\cite{BdJ00}) suggest that high-surface brightness
spirals are generally close to maximal disks.

\begin{figure*}[t]
\begin{center}
\resizebox{\figwidth}{!}{\includegraphics{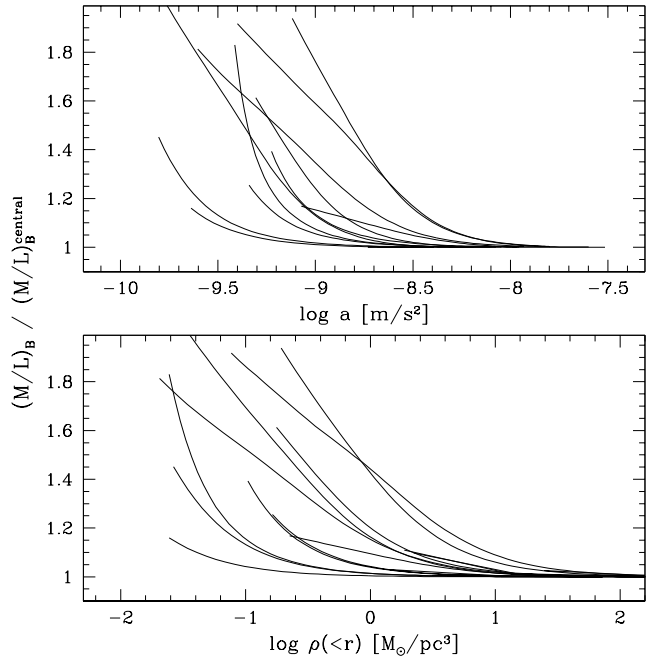}}
\caption[dsb]{\small Upper panel: B-band $M/L_B$, normalized by the
central $M/L_B$, versus the gravitational acceleration $v^2/r$ for all
galaxies from the EK subsample.  The ``bending upwards'' occurs at
accelerations $a\sim 10^{-9} {\rm m\, sec}^{-2}$, a factor of 10 higher
than typical in spiral galaxies.  Lower panel: Normalized $M/L_B$
versus mean interior density. }
\label{accdensml}
\end{center}
\end{figure*}

One possible explanation for the much larger densities and probably
phase-space densities of elliptical galaxy halos might be that some of
the dark matter inferred in the inner regions of elliptical galaxies
is baryonic. The evidence for large $M/L$-ratios in ellipticals at
large radii, of order one hundred, from both X-ray (e.g., Matsushita
\etal \cite{Metal98}) and weak lensing data (Griffiths \etal
\cite{Gri98}, Wilson \etal \cite{WKLC00}), together with the lack of
microlensing towards the LMC in the Milky Way (Alcock \etal
\cite{Alc+00}), does however not make this an attractive explanation for
galactic halos in their entirety; a separate inner baryonic dark matter
component would be needed.

An alternative possibility is that most elliptical galaxies formed at
high redshift from progenitors with higher densities than seen in
present-day spiral galaxies.  If halo core densities are proportional
to virial densities, which in turn depend on the density of the
Universe at the time of collapse, then the result above implies that
elliptical galaxy halos have collapsed at redshifts $z_E \gta 25^{1/3}
(1+z_S) -1$, i.e., $z_E\gta 5$ if the halos of spiral galaxies of
similar luminosity formed at redshifts $z_S\gta1$. Thus our result may
indicate that giant elliptical galaxies are old, consistent with
evidence from the fundamental plane (van Dokkum \& Franx \cite{vDF96},
van Dokkum \etal \cite{vD+98}, Bender \etal \cite{Ben+98}) and
line-strength indices (Bender, Ziegler \& Bruzual \cite{Ben+96}).
Unfortunately, while this argument is plausible, it is not conclusive
until the relation between the shallow central halo profiles inferred
from observations and the cuspy halos predicted from hierarchical
collapse of dark matter (Navarro, Frenk \& White \cite{NFW96}) is
understood.  Moreover, it would appear to be at odds with the
observation of substantial merging in a moderate z=0.83 cluster (van
Dokkum \etal \cite{vD+99}), unless the progenitors had unusually high
halo densities also, and also with the expectation that some
elliptical galaxies should have formed recently from mergers of normal
spiral galaxies (e.g., Schweizer \cite{Schw98}).

The plot of halo core radius against luminosity displays considerably
larger scatter for both subsamples than the plot of $R_e$ versus $L$
and consequently also the inferred halo density shows considerable
scatter at a given luminosity.  What is the origin of this increased
scatter? First, it is possible that modelling uncertainties contribute
to the larger scatter in the derived $r_{c,h}$. However, we do not
think that this can be the whole explanation. The three galaxies at
the upper boundary of the points in the $r_{c,h}$-$L$-plot which have
the best kinematic data for their kind, NGC 1399, NGC 3379 and NGC
6703, all have $\log(r_{ch}/R_e) =(0.5:0.6)$ and show evidence for
small if any amount of dark matter within the modelled range. On the
other hand, the best-determined galaxies near the lower boundary with
the best evidence for additional dark matter, NGC 2434, 7507, 7626,
have $\log(r_{ch}/R_e) =(-0.15:-0.35)$.  Fig.~\ref{vcCurves1} shows
that the two groups have significantly different CVC shapes {\sl for
  the visible component only}: for the first group the visible CVC is
almost coincident with the dynamically inferred total CVC to $1R_e$,
whereas for the second group there are significant mass discrepancies
already at $1R_e$. Thus elliptical galaxies at fixed $L_B$ appear to
have a range of luminous matter CVCs and hence dark matter CVCs, even
though the total CV rotation curves are fairly similar. Also, while
the most rapidly rotating galaxies in the sample, for which we would
expect the largest systematic errors in the modelling, have
predominantly positive residuals with respect to the least-square
line, the non-rotating galaxies populate the entire distribution of
residuals including the extremes.  Thus we believe that most of the
scatter in the inferred $\log(r_{ch}/R_e)$ is not due to modelling
effects, but reflects physical differences between the sample
galaxies.

Could the amount of dark mass in the centers of preferentially those
galaxies with apparently large $r_{ch}$ have been underestimated with
our minimum halo models? Then we would expect that the derived central
$M/L_B$ values of these galaxies should be systematically high for
their luminosity, i.e., we would expect a correlation of positive
residuals from the $M/L_B-L$ relation in Fig.~\ref{lmllum} with
positive residuals in $r_{ch}-L$. However, the $M/L_B$ residuals for
galaxies with large and small halo core radii for their luminosities
do not show a systematic difference in the present sample.

Thus, we believe the most likely explanation for the larger scatter in
$r_{ch}$ is that galaxies of similar luminosity have different dark
matter core radii and central densitites determined by the particulars
of the merging process in which they were made. In this case the
scatter in inferred halo density (an order of magnitude or more)
should perhaps reflect mainly the halo densities of the progenitors at the
time of formation, with the highest (lowest) densities corresponding
to the earliest (latest) mergers.  We have attempted to test this by
plotting the minimum halo densities from Fig.~\ref{rcLum} versus the
population ages of Section~\ref{secpop}, but no convincing correlation
is seen in the present data. 

Figure \ref{accdensml} finally shows mass-to-light ratios $M/L_B$
normalized by the central value, as a function of acceleration $a =
v_c^2(r)/r$ and mean interior density $\rho(<r)\equiv 3v_c^2(r)/4\pi G
r^2$, for all galaxies in the EK subsample with extended data.  The
estimated uncertainty in $a$ from that in $v_c$ and distance is of
order $50\%$. Figure \ref{accdensml} shows that the ``bending
upwards'' which indicates the onset of dark matter takes place at
systematically higher accelerations $a$ than in spiral galaxies
(McGaugh \cite{McG99}), by about one order of magnitude, which is a
consequence of the higher halo mass densities in elliptical galaxies.
This suggests that the acceleration scale found in spiral galaxies is
not universal as required by the modified gravity theory MOND (Milgrom
\cite{Mil83}), i.e., that in elliptical galaxies an additional
gradient in $M/L$ would be required besides MOND.


\section{Conclusions}

Based on a uniform dynamical analysis of photometric and line-profile
shape data for 21 mostly luminous, slowly rotating, and nearly round
(17 E0/E1, 4 E2) elliptical galaxies by Kronawitter \etal
(\cite{K+00}), we have investigated the dynamical family relations and
dark halo properties of ellipticals. Our main results are as follows:

(1) The circular velocity curves (CVCs) of elliptical galaxies are
flat to within $\simeq 10\%$ for $R\gta 0.2R_e$ to at least $R\gta
2R_e$, independent of luminosity.  This argues against strong
luminosity segregation in the dark halo potential.

(2) Most ellipticals are moderately radially anisotropic, with average
$\beta\simeq 0-0.35$, again independent of luminosity.

(3) The dynamical structure of ellipticals is surprisingly uniform.
The maximum circular velocity is accurately predicted by a suitably
defined central velocity dispersion. 

(4) Elliptical galaxies follow a Tully-Fisher (TF) relation with marginally
shallower slope than spiral galaxies. At given circular velocity, they
are about 1 mag fainter in B and about 0.6 mag in R, and appear to have
slightly lower baryonic mass than spirals, even for the maximum
$M/L_B$ allowed by the kinematics.

(5) The residuals from the TF and Fundamental Plane (FP) relations do not
correlate with dynamical anisotropy $\beta$.

(6) The luminosity dependence of $M/L$ indicated by the tilt of the FP
corresponds to a real dependence of dynamical $M/L$ on $L$. The tilt
of the FP is therefore not due to deviations from homology or a
variation of dynamical anisotropy with $L$, although the slope of
$M/L$ versus $L$ could still be influenced by photometric
non-homology.  The tilt can also not be due to an increasing dark matter
fraction with $L$, unless (i) the most luminous ellipticals have a
factor $>3$ less baryonic mass than spiral galaxies of the same
circular velocity, (ii) the range of IMF is larger than currently
discussed, and (iii) the IMF or some other population parameter
varies systematically along the luminosity sequence such as to undo
the increase of $M/L$ expected from simple stellar population models
for more metal-rich luminous galaxies. This seems highly unlikely.

(7) The tilt of the FP is therefore best explained as a stellar
population effect.  Population models show that the values and the
change with $L_B$ of the maximal dynamical $M/L_B$s are consistent
with the stellar population $M/L_B$s based on published metallicities
and ages within the uncertainties of IMF and distance scale.  The main
driver is therefore probably metallicity, and a secondary population
effect is needed to explain the K-band tilt.

(8) The population models show that we would have underestimated
the luminous masses by as much as a factor $\approx 2$ only if (i) the
flattest IMFs at low stellar masses discussed for the Milky Way are
applicable to our elliptical galaxies, and simultaneously (ii) a short
distance scale ($H_0\simeq 80\kms\mpc^{-1}$) turns out to be correct.  For
lower values of $H_0$ and/or the other IMFs investigated in Section
\ref{secpop} the difference is smaller. Together with (4) this makes
it likely that elliptical galaxies have indeed nearly maximal $M/L_B$
ratios (minimal halos).

(9) Despite the uniformly flat CVCs, there is a spread in the ratio of
the CVCs from luminous and dark matter, i.e., in the radial variations
of cumulative mass-to-light ratio. The sample includes galaxies with
no indication for dark matter within $2R_e$, and others where the best
dynamical models result in local $M/L_B$s of 20-30 at $2R_e$.  As in
spiral galaxies, the combined rotation curve of the luminous and dark
matter is flatter than those for the individual components
(``conspiracy'').

(10) In models with maximum stellar mass, the dark matter contributes
$\sim 10-40\%$ of the mass within $R_e$. Our flat rotation curve
models, when extrapolated beyond the range of kinematic data, predict
equal interior mass of dark and luminous matter at $\sim 2-4R_e$,
consistent with results from the X-ray temperature - velocity
relation.

(11) Even in these maximum stellar mass models, the halo core
densities and phase-space densities are at least $\sim 25$ times
larger and the halo core radii $\sim 4$ times smaller than in spiral
galaxies of the same circular velocity. Correspondingly, the increase
in $M/L$ sets in at $\sim 10$ times larger acceleration than in
spirals.  This could imply that elliptical galaxy halos collapsed at
redshifts $z>5$ or that some of the dark matter in ellipticals might
be baryonic.

\begin{acknowledgements} 
  We thank D.~Forbes for providing ages and metallicities for some of
  our sample galaxies in electronic form, C.~Maraston for help with
  her stellar population models, and M.~Samland for assistance with
  IDL. We acknowledge helpful discussions with H.-W.~Rix and
  G.A.~Tammann.  OG and AK were supported by grant 20-56888.99 from
  the Schweizerischer Nationalfonds, RPS and RB acknowledge the
  support by DFG grant SFB 375.
\end{acknowledgements}

\label{lastpage}

\end{document}